\documentclass{amsart}

\usepackage{amsmath}
\usepackage{mathtools}
\usepackage{algorithm}
\usepackage{graphicx}
\usepackage{algpseudocode}
\usepackage[utf8]{inputenc}
\usepackage[T1]{fontenc}
\usepackage{url}
\usepackage[binary-units=true,per-mode=symbol]{siunitx}
\usepackage{paralist}
\usepackage{footnote}
\usepackage{caption}
\usepackage{subcaption}
\usepackage[bottom]{footmisc}
\usepackage{array}
\usepackage[shortlabels]{enumitem}
\usepackage{acronym}
\usepackage{nomencl}
\usepackage{amsthm}
\usepackage{thmtools}
\usepackage{nicefrac}
\usepackage{hyperref}
\usepackage{xpatch}
\usepackage{balance}

\def\baselinestretch{1}

\newtheoremstyle{definition}
{0.15\baselineskip}
{0.15\baselineskip}
{}
{}
{\bfseries}
{:}
{.5em}
{}

\newtheoremstyle{theorem}
{0.15\baselineskip}
{0.15\baselineskip}
{}
{}
{\bfseries}
{:}
{.5em}
{}

\newtheoremstyle{subdefinition}
{0.15\baselineskip}
{0.15\baselineskip}
{}
{}
{\bfseries}
{:}
{.5em}
{}

\newtheoremstyle{lemma}
{0.15\baselineskip}
{0.15\baselineskip}
{}
{}
{\bfseries}
{:}
{.5em}
{}

\declaretheoremstyle[
spaceabove=0pt,%
spacebelow=4pt,%
headfont=\normalfont\itshape,%
postheadspace=0.5em,%
qed=\qedsymbol%
]{mystyle} 
\declaretheorem[name={Proof},style=mystyle,unnumbered,
]{prf}

\AtBeginDocument{%
	\addtolength\abovedisplayskip{-0.2\baselineskip}%
	\addtolength\belowdisplayskip{-0.2\baselineskip}%
}

\newcommand{\imgWidth}{0.7\linewidth}

\theoremstyle{definition}

\newtheorem{definition}{Definition}[section]

\theoremstyle{theorem}

\newtheorem{theorem}{Theorem}[section]
\newtheorem{lemma}[theorem]{Lemma}

\newcommand{\pseudoCaptionWidth}{-0.25cm}

\makenomenclature

\xpatchcmd{\thenomenclature}{%
	\section*{\nomname}
}{
}{\typeout{Success}}{\typeout{Failure}}

\begin{document}
		\textbf{© ACM 2019. This is the author's version of the work. It is posted here for your personal use. Not for redistribution. The definitive Version of Record was published in ACM SIGCOMM 2019, 
		\url{http://dx.doi.org/10.1145/3341302.3342067}}
		\\
		\\
	\begin{abstract}
		Mobile devices apply neighbor discovery (ND) protocols to wirelessly initiate a first contact within the shortest possible amount of time and with minimal energy consumption. For this purpose, over the last decade, a vast number of ND protocols have been proposed, which have progressively reduced the relation between the time within which discovery is guaranteed and the energy consumption. In spite of the simplicity of the problem statement, even after more than 10 years of research on this specific topic, new solutions are still proposed even today.
Despite the large number of known ND protocols, given an energy budget, what is the best achievable latency still remains unclear. 
This paper addresses this question and for the first time presents safe and tight, duty-cycle-dependent bounds on the worst-case discovery latency that no ND protocol can beat. 
Surprisingly, several existing protocols are indeed optimal, which has not been known until now. We conclude that there is no further potential to improve the relation between latency and duty-cycle, but future ND protocols can improve their robustness against beacon collisions.
	\end{abstract}
	\title{On Optimal Neighbor Discovery}
	\author{Philipp H. Kindt and Samarjit Chakraborty}
	\thanks{The authors are with the Chair of Real-Time Computer Systems (RCS), Department of Electrical and Computer Engineering, Technical University of Munich (TUM), Germany. Email: philipp.kindt/samarjit[at]tum.de.} 
	\maketitle
	\hypersetup{
		pdftitle = {On Optimal Neighbor Discovery},
		pdfauthor = {Anonymous Authors}
	}
	
	\section{Introduction}

Wireless networks that operate without any fixed infrastructure are rapidly growing in importance. Since all devices in such \acp{MANET} run on batteries or rely on intermittently available energy-harvesting sources, the energy spent for communication needs to be as low as possible. Typically, \ac{MANET} radios are duty-cycled and wake up only for short durations of time for carrying out the necessary communication and then go back to a sleep mode. While such duty-cycled communication schemes are easy to realize when the clocks of all devices are synchronized and their wakeup schedules are known by all participants of the network, asynchronous communication (i.e., communication without synchronized clocks) remains a challenging problem.
One of the most important asynchronous procedures is establishing the first contact between different wireless devices, which is referred to as \textit{\ac{ND}}. \\
 
\vspace{\pseudoCaptionWidth}
\noindent\textbf{Neighbor Discovery:} \ac{ND} is used by a device for detecting other devices in range.
This could be for clock synchronization and establishing a connection, after which more data can be exchanged in a synchronous fashion.
Efficient \ac{ND} is characterized by achieving the shortest possible discovery latency for a given energy budget. Towards this, a large number of \ac{ND} protocols have been proposed till date, see ~\cite{mcglynn:02,margolies:16,you:11,vasudevan:2009, schurgers:02,Tseng:02,dutta:08,Kandhalu:10,bakht:12,meng:14, meng:16,chen:15,zhang:12,Sun:14,qiu:16,julien:17,kindt:17a,kindt:18, wei:16, zheng:03, zheng:06, chen:18, kandhalu:13, lee:19,kim:13,vasudevan:13,liu:11,vasudevan:05,jakllari:07,zhang:08,karowski:11,zeng:11,borbash:16,wang_2:13,zhang:17,guo:17,zhang:17_b,cao:18,chen:17,chen:16_b,lee:15,yang:15,wang:14,hess:14,zhang:13,li:13,cohen:11,yang:09}.
Among these, ~\cite{schurgers:02,Tseng:02,dutta:08,Kandhalu:10,bakht:12,meng:14, meng:16,chen:15,zhang:12,Sun:14,qiu:16,julien:17,kindt:17a,kindt:18} for example, concern \textit{deterministic} discovery. Here, given the protocol parameters, an upper bound on the discovery latency can be determined. The problem of \textit{pairwise discovery} between two devices is of fundamental importance, since in many scenarios, devices join the network gradually and only a master device and the newly joining one carry out the discovery procedure simultaneously. Moreover, the process of discovering multiple devices always relies on pairwise \ac{ND}.

Over the years, successive \ac{ND} protocols have improved their discovery latencies for given energy budgets.
For example, the \textit{Griassdi}~\cite{kindt:17a} protocol proposed in 2017 claims to achieve by $87 \%$ lower worst-case latencies than \textit{Searchlight-Striped}~\cite{bakht:12} that was proposed in 2012. 
However, despite the significant attention the \ac{ND} problem has received over the past $15+$ years, the fundamental question of what is the theoretically lowest possible discovery latency that any \ac{ND} protocol could guarantee for a given energy budget still remains unanswered.
\\

\vspace{\pseudoCaptionWidth}
\noindent\textbf{Performance of ND Protocols:}
In the absence of such a bound, the performance evaluations of different \ac{ND} protocols have often been very subjective.
The results of such evaluations relied on the choice of protocols, their parametrizations and the assumed setups. Hence, while a certain protocol might outperform others in such a comparison, it might perform differently if the parametrization or setup is changed.
In addition, most known protocols, e.g., \cite{dutta:08, bakht:12, Sun:14}, subdivide time into multiple slots and are hence referred to as \textit{slotted}. The device sleeps in most slots, whereas some slots are active and used for communication. In each active slot, a device sends a beacon at the beginning and/or end of the slot and listens for incoming beacons in the meantime. Discovery occurs once two active slots overlap in time. Here, performance is quantified in terms of the worst-case number of slots until discovery is guaranteed. Though a certain protocol could perform better than another in terms of the number of slots, such comparisons are heavily dependent on the supported range of slot lengths. As a result, such comparisons in terms of slots and not directly in terms of time are often not meaningful. 
Moreover, despite slotted protocols having been studied thoroughly in the literature, many protocols that are frequently used in practice, e.g., \ac{BLE}, do not rely on a slotted paradigm. They schedule reception windows and beacon transmissions with periodic intervals and offer three degrees of freedom that can be configured freely (viz., the periods for reception and transmission, and the length of the reception window). The high practical relevance of such \textit{\ac{PI}-based} protocols is underpinned by the 4.7 billion \ac{BLE}-devices that were expected to be sold in 2018~\cite{statista_BLE:2018}. 
It has recently been shown that the parametrizations for \ac{ND} in \ac{BLE} networks proposed by official specifications~\cite{BleFindMeProf} lead to performances far from the optimum~\cite{kindt:18}.
This has raised the interest to fully understand such \textit{slotless} \ac{ND} procedures. In particular, finding beneficial parametrizations for periodic interval-based protocols has been studied in the literature recently, e.g., in~\cite{kindt:18, julien:17, kindt:17a}. However, until today, it is neither clear whether the proposed parametrizations are actually optimal, nor how such protocols compare to the slotted ones in terms of performance.
In summary, despite the large volume of available literature, it is not possible to meaningfully assess and classify the performance of \ac{ND} protocols in a purely objective fashion. \\

\vspace{\pseudoCaptionWidth}
\noindent \textbf{This Paper:}
In this paper, we study the fundamental limits of pairwise, deterministic \ac{ND}. In particular, we establish a relationship between the optimal discovery latency, channel utilization (and hence beacon collision rate) and duty-cycle. No pairwise \ac{ND} protocol can achieve lower discovery latencies than the ones established in this paper. The resulting bounds not only give important insights into the design of \ac{ND} protocols, but will serve as a baseline for more objective performance comparisons. Surprisingly, our analysis shows that some recently proposed protocols actually perform optimally and cover the entire discovery latency/channel utilization/duty-cycle Pareto front. The optimality results of such protocols were not known until now. In particular, the coverage of the Pareto front implies that there is no further potential for improvement. However, there is still potential to improve the robustness against beacon collisions, which might occur frequently when many devices carry out \ac{ND} simultaneously.\\

\vspace{\pseudoCaptionWidth}
\noindent\textbf{Principle of \ac{ND}:} 
In general, a radio can either be in a sleep state, listen to the channel or transmit a beacon. Hence, the basic building blocks of a \ac{ND} protocol are given by these three operations and any \ac{ND} protocol can be represented as a unique pattern of them. For a higher power-budget, the number of beacons and/or the number or lengths of reception windows can be increased and a discovery procedure is successful once a beacon overlaps with a reception window on another device.
Since the design space of all possible reception and transmission patterns allows for an infinite number of possible configurations, determining the optimal pattern and its performance through any form of exhaustive search or numerical method is not possible. Further, as outlined above, most work on \ac{ND} has focused on slotted protocols and therefore studied only a small part of the design space. As a result, the problem of assessing the optimal performance of \ac{ND} has so far remained unsolved.
\\


\vspace{\pseudoCaptionWidth}
\noindent\textbf{\ac{ND} Scenarios:} For different scenarios, the \ac{ND} problem appears in different forms, and we provide bounds on the discovery latency for many of them. 
First, it is obvious that if two devices \textit{E} and \textit{F} both have the same beacon and reception patterns, their discovery properties are \textit{symmetric}. This implies that device~\textit{E} discovers device~\textit{F} with the same worst-case latency for a given duty-cycle as \textit{F} discovering \textit{E}. Several publications, e.g., \cite{zheng:06, dutta:08, kindt:18}, have studied this special case of \textit{symmetric} duty-cycles, for which we present a bound on the discovery latency. If both devices run different patterns (for example, due to different duty-cycles), the discovery properties are \textit{asymmetric}. For the asymmetric case, we provide a bound on the discovery latency when each device is aware of the other device's configuration.

Another important question we answer in this paper is the partitioning of the duty-cycle, which corresponds to the energy-budget of a device. The duty-cycle of a device is the fraction of time it is active. On the other hand, channel utilization is the fraction of time a device occupies the channel, which is between zero and its duty-cycle. Beacon collision rates are solely determined by the channel utilizations of the devices in range.
For the case when the channel-utilization (and hence collision rate) is unconstrained, we derive the ratio between transmission and reception times that minimizes the discovery latency. 

In the case of many devices discovering each other, the channel utilization of each device has to be constrained for limiting the collision rate.
In this paper, we therefore not only derive bounds for the discovery latency that any protocol can guarantee for a given duty-cycle, but also for the case where both duty-cycle and the maximum channel-utilization are provided.
To the best of our knowledge, no such protocol-agnostic bounds on discovery latency for the \ac{ND} problem has been derived until now. In particular, this paper makes the following contributions.\\

\vspace{\pseudoCaptionWidth}
\noindent 
\textbf{Technical Contributions:}
We present the following bounds on the discovery latency of deterministic \ac{ND} protocols.
\begin{compactenum}
	\item The lowest discovery latency \textit{any} symmetric and asymmetric pairwise \ac{ND} protocol can guarantee for a given duty-cycle and hence energy consumption.
	Recall that in symmetric \ac{ND}, all devices operate using the same duty-cycle, whereas in asymmetric \ac{ND} devices use different duty-cycles. 
	\item A discovery latency bound for the case where the channel utilization is additionally constrained.
	\item Bounds for the following three cases where two devices \textit{E} and \textit{F} discover each other.
	(a)~Only \textit{E} needs to discover \textit{F}, whereas \textit{F} does not need to discover \textit{E}. (b)~Either \textit{E} discovers \textit{F} or \textit{F} discovers \textit{E}, but both discovering each other is not possible. (c)~Both \textit{E} and \textit{F} mutually discover each other.
\end{compactenum}

We further study the relation between our bounds and previously known ones \cite{zheng:03, zheng:06, meng:14, meng:16}, which are all limited to slotted protocols. These bounds are given in terms of a worst-case number of slots until discovery is guaranteed, where the discovery latency also depends on the slot length.
However, how small a slot length can be is difficult to answer, while it is known that slot lengths cannot be made arbitrarily small. Therefore, discovery latencies in terms of time have not been derived, which we address in this paper.
Finally, while most previous work has focused on slotted protocols, we show that when channel utilization is unconstrained, only slotless protocols can perform optimally, whereas slotted ones cannot. This result is important because in many IoT scenarios devices join gradually and only a pair of devices participate in \ac{ND} at any point in time. Here, channel utilization is therefore not of concern. \\

\vspace{\pseudoCaptionWidth}
\noindent\textbf{Organization of the paper:}
The rest of this paper is organized as follows. In Section~\ref{sec:related_work}, we present related work on discovery latency bounds of \ac{ND} protocols. Next, in Section~\ref{sec:nd_protocols}, we provide a formal description of a generic \ac{ND} procedure. Based on this, in Section~\ref{sec:analysis}, we derive a list of properties that deterministic \ac{ND} protocols need to guarantee. Recall that deterministic \ac{ND} protocols are ones for which bounded discovery latencies can be guaranteed.
We derive such latency bounds in Section~\ref{sec:bounds}. Finally, in Section~\ref{sec:previously_known_bounds}, we relate the latency bounds of multiple existing \ac{ND} protocols to the bounds obtained in this paper. 
Throughout this paper, we make a couple of simplifying assumptions. These assumptions are only for the ease of exposition and are relaxed in Section~\ref{sec:validity_of_assumptions}.
A table of symbols and additional proofs are given in the appendix.

\section{Related Work}
\label{sec:related_work}
In this section, we describe existing efforts to determine bounds on the discovery latency that any \ac{ND} protocol can achieve, and relate them to this paper.\\

\vspace{\pseudoCaptionWidth}
\noindent
\textbf{Bounds for Slotted Protocols:}
As discussed above, the vast majority of \ac{ND} protocols proposed in the literature follow a slotted paradigm, in which reception and transmission are temporally coupled into slots. A bound on the discovery latency of slotted protocols has been studied in \cite{zheng:03, zheng:06}. 
Here, it has been shown that for guaranteeing discovery within $T$ slots, every device needs to have at least $k=\sqrt{T}$ active slots. Therefore, if e.g., $k=2$ out of $T = 4$ slots are active, then discovery can be guaranteed within four slots with a duty-cycle of $50\%$, whereas if $k = 4$ and $T = 16$, discovery can be guaranteed within 16 slots with a duty-cycle of $25\%$. Determining the schedule of active slots that realizes this bound relies on \textit{cyclic difference sets}~\cite{zheng:03}. Since only a very limited number of such difference sets are known, slotted protocols utilizing this bound can only be realized for a few duty-cycles that correspond to these known difference sets. Subsequently proposed protocols, such as Disco~\cite{dutta:08}, Searchlight~\cite{bakht:12} and U-Connect~\cite{Kandhalu:10} for the same discovery latency require more active slots than defined by this bound. But they are more flexible in terms of duty-cycles they can realize.
Other recent work \cite{meng:14, meng:16} claims to have superseded this bound. By sending an additional beacon outside the slot boundaries in a schedule defined by difference sets, a tighter bound than described in \cite{zheng:03, zheng:06} can be reached.

Being on slotted protocols, the bounds in \cite{zheng:03, zheng:06, meng:14, meng:16} are all given in terms of a worst-case number of slots within which discovery is guaranteed. The corresponding bounds in terms of time depend on the slot length $I$. 
The minimum size of such a slot, among other factors, also depends on the hardware, and cannot be made arbitrarily small.
Consequently, no bounds on the discovery latency in terms of time for slotted protocols have been known until now. 
This issue is addressed later in this paper.\\

\vspace{\pseudoCaptionWidth}
\noindent
\textbf{Bounds for PI-based Protocols:}
Given a tuple of parameter values $(T_a, T_s, d_s)$, a method to compute the worst-case discovery latency for \ac{PI}-based protocols was provided in \cite{kindt:15}. However, since there are infinite numbers of possible parametrizations $(T_a, T_s, d_s)$, and because of the computation scheme provided in \cite{kindt:15}, which parametrization leads to the lowest discovery latency has so far remained unknown. 
Recently, \cite{kindt:18} and \cite{kindt:17a} proposed parametrization schemes that can compute parameters $(T_a, T_s, d_s)$ to realize any given duty-cycle. However, the optimality of such parametrizations in terms of discovery latency has not been established.\\

\vspace{\pseudoCaptionWidth}
\noindent
\textbf{Generic Approaches:}
Unlike the work described above that was specific to slotted or \ac{PI}-based protocols, protocol-agnostic bounds were presented in \cite{bradonjic:12, barenboim:14}. 
In particular, they give an asymptotic latency bound in the form of $\Theta(d)$, where $d$ is ``the discretized uncertainty period of the clock shift between the two processors''~\cite{bradonjic:12}. Hence, this bound depends on the degree of asynchrony between the clocks of a sender and a receiver. First, the asymptotic nature of such a bound is very different from the concrete time bounds that have been pursued within the computer communications community, e.g., \cite{zheng:03,zheng:06,meng:14,meng:16}, and the ones presented by us in this paper. Second, this community has also focused on bounds that depend on duty-cycle and hence energy budget, which are of direct practical relevance.
For these reasons, the bounds from \cite{bradonjic:12, barenboim:14} are not comparable to those that have been more commonly pursued, and also to those presented in this paper.
\section{Neighbor Discovery Protocols}
\label{sec:nd_protocols}
\subsection{Definition}
\label{sec:definition_of_nd_protocols}
In this section, we formally define the \ac{ND} procedure and its associated properties.

\begin{figure}[tb]
	\begin{subfigure}[b]{\imgWidth}
		\centering
		\includegraphics[width=\columnwidth]{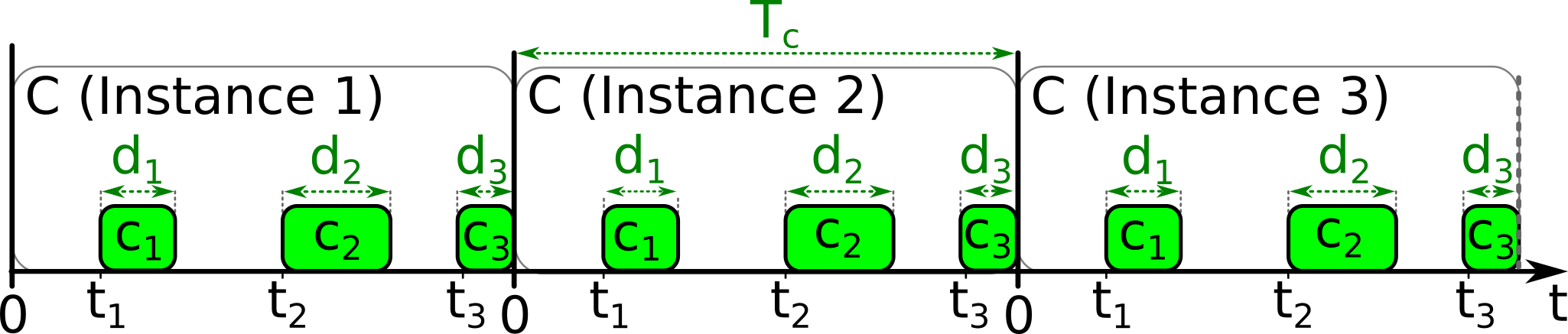}
		\caption{Reception window sequence. Here, $\mathbf{C = (c_1, c_2, c_3)}$.}
		\label{fig:nd_protocol_rcv} 
	\end{subfigure}
	\begin{subfigure}[b]{\imgWidth}
		\centering
		\includegraphics[width=\columnwidth]{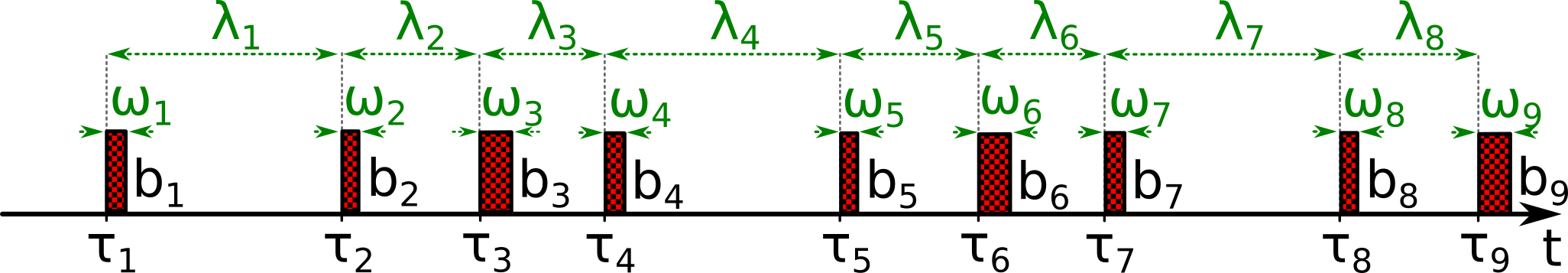}
		\caption{Beacon sequence $B =b_1, b_2, ..., b_9$.}
		\label{fig:nd_protocol_beacons} 
	\end{subfigure}
	\caption{Reception window and beacon sequences.}
	\label{fig:nd_protocol_all} 
\end{figure}

\begin{definition}[Reception Window Sequence]
	\label{def:chunk}
	Let the time windows during which a device listens to the channel be given by the tuples $c_1 = (t_1, d_1), c_2 = (t_2, d_2), c_3 = (t_3, d_3),...$, where each \textit{reception window} $c_i$ starts at time $t_i$ and ends $d_i$ time-units later (see Figure \ref{fig:nd_protocol_rcv}). A \textit{reception window sequence} $C = c_1, c_2, ..., c_{n}$ could be of finite or infinite length. In this paper, for simplicity of notation, we refer to such finite length sequences by $C$ and infinite length sequences by $C_\infty$. 
\end{definition}
\nomenclature{$t_{i}$}{Point in time the reception window $i$ begins at}
\nomenclature{$d_{i}$}{Time duration of reception window $i$}
\nomenclature{$c_i$}{Reception window $i$}
\nomenclature{$C/C_\infty$}{Finite/infinite reception window sequence}
For the simplicity of exposition, throughout this paper, we always assume that any $C_\infty$ is an infinite concatenation of some finite length sequence $C$. For such $C_\infty$, we define $n_C = |C|$ (i.e., the number of windows contained in $C$). Further, we denote the time between the ends of two consecutive instances of $C$ as the \textit{reception period} $T_C$. It is worth mentioning that all our bounds remain valid also for sequences $C_\infty$ that are not given by concatenating the same C, as we show in Appendix~\ref{app:non_repetitive_beacon_sequences}.
\nomenclature{$n_C$}{Number of reception windows contained in a finite length reception window sequence $C$, whose concatenations form an infinite sequence $C_\infty$}
\nomenclature{$T_C$}{Time between the ends of two consecutive instances of the finite reception window sequence $C$, whose concatenations form an infinite sequence $C_\infty$}
We assign a time-axis to every instance of $C$. For convenience, which will become clear later, the origin of time in a certain instance of $C$ will start at the end of the last reception window of the previous instance, as depicted in Figure~\ref{fig:nd_protocol_rcv}. In this figure, $C$ consists of three reception windows (i.e., $c_1, c_2, c_3$), and three concatenated instances of $C$ are shown. For example, the origin of the time-axis for Instance~2 lies at the end of $c_3$ in Instance~1.

\begin{definition}[Beacon Sequence]
\label{def:beaconing_schedule}
A sequence of beacons $B = b_1, b_2,...,b_{m}$ sent at the time-instances $\tau_{1}, \tau_{2},...,\tau_{m}$, as depicted in Figure~\ref{fig:nd_protocol_beacons}, is called a \textit{beacon sequence} of length $m$. The transmission durations of these beacons are given by $\omega_{1}, \omega_{2},...,\omega_{m}$. A sequence of infinite length (i.e., $m \to \infty$) is denoted by $B_\infty$. 
\nomenclature{$b_i$}{Beacon $i$}
\nomenclature{$B/B_\infty$}{Finite/infinite beacon sequence}
\nomenclature{$\omega_{i}$}{Transmission duration of beacon $i$}
\nomenclature{$\tau_{i}$}{Time at which beacon $i$ is sent}
\nomenclature{$T_B$}{Time-period of a repetitive, infinite beacon sequence}
\nomenclature{$m_B$}{Period of a repetitive beacon sequence (in terms of number of beacons)}

We denote infinite length beacon sequences $B_\infty$ that are given by concatenations of a finite beacon sequence $B$ as \textit{repetitive beacon sequences}. In such repetitive sequences, $m_B = |B|$ and the time between the endings of two consecutive instances of $B$ is given by $T_B$. Unlike for reception window sequences, we do not restrict ourselves to repetitive infinite beacon sequences. However, we will prove that all beacon sequences that optimize the relevant metrics of a \ac{ND} procedure are repetitive when the corresponding reception window sequence is also repetitive. 

We indicate an arbitrary shorter sequence $B^\prime$ to be a part of a longer sequence $B$ by using the notation $B^\prime \in B$. For example, in Figure~\ref{fig:nd_protocol_beacons}, $B^\prime = b_2, b_3, b_4, b_5, b_6\in B$. Further, the time between the beginnings of beacon $b_i$ and beacon $b_{i+1}$ is called the \textit{beacon gap} $\lambda_i$. It is $\lambda_{i} = \tau_{i+1}-\tau_{i}$.
\nomenclature{$\lambda_{i}$}{Gap between beacon $i$ and beacon $i+1$}

\end{definition}
\begin{definition}[\ac{ND} Protocol]
	\label{def:nd_protocol}
	A tuple of an infinite beacon and reception window sequence $(B_\infty,C_\infty)$ is called a \textit{\ac{ND}} protocol. 
\end{definition}

In this paper, unless explicitly stated, we assume that $C_\infty$ and $B_\infty$ stem from two different devices \textit{E} and \textit{F}. When it is necessary to explicitly specify the device that a sequence is scheduled on, we use the notation $C_{E,\infty}$ or $B_{F,\infty}$, where \textit{E} and \textit{F} refer to device \textit{E} or \textit{F} respectively. We also apply this notation to reception windows and beacons, e.g., $b_{E,1}$ refers to beacon~1 on device~\textit{E} and $c_{F,1}$ refers to reception window~1 on device~\textit{F}.

The most important properties of a \ac{ND} protocol are its worst-case latency $L$, its duty-cycle $\eta$, and its channel utilization $\beta$, as defined next.
\begin{definition}[Worst-Case Latency]
\label{def:wc_latency}
Given two devices \textit{E} and \textit{F}, where \textit{E} runs an infinite beacon sequence and \textit{F} an infinite reception window sequence, the \textit{worst-case latency} $L$ is the earliest possible time after which an overlap of a beacon from \textit{E} with a reception window of \textit{F} is guaranteed, measured from the point in time both devices come into the range of reception.
\end{definition}
\nomenclature{$L$}{Worst-case latency}

\begin{definition}[Duty-Cycle]
	\label{def:duty_cycle}
	The \textit{transmission duty-cycle} $\beta$ of a device is the fraction of time it spends for transmission, whereas the \textit{reception duty-cycle} $\gamma$ is the fraction of time spent for reception. In general, depending on the configuration of the radio (e.g., transmit power and receiver gain), transmission incurs a different power consumption than reception. Therefore, the total \textit{duty-cycle} $\eta$ is given as a weighted sum $\eta = \gamma + \alpha \beta$, where the weight $\alpha$ is the ratio of transmission and reception powers, i.e., $\alpha = \nicefrac{P_{Tx}}{P_{Rx}}$.  
	For a radio running a tuple of sequences $(B_\infty, C_\infty)$, it is:
	\begin{equation}
	\label{eq:duty_cycle_def}
	 \quad \beta = \lim_{m \to \infty} \frac{\sum_{i = 1}^{m-1} \omega_i}{\tau_m - \tau_1}, \quad \gamma = \lim_{n \to \infty} \frac{\sum_{i = 1}^{n-1} d_i}{t_n - t_1}, \quad \eta = \alpha \beta + \gamma
	\end{equation}
\end{definition}
\nomenclature{$\eta$}{Duty-cycle}
\nomenclature{$\beta$}{Duty-cycle for transmission, which is equivalent to the channel utilization}
\nomenclature{$\gamma$}{Duty-cycle for reception}
\nomenclature{$\alpha$}{Ratio of the power spent for transmission over the power spent for reception}
\nomenclature{$P_{Tx}, P_{Rx}$}{Power consumption of a radio for transmission or reception, respectively}
The transmission duty-cycle $\beta$ is the same as the \textit{channel utilization}.
The duty-cycle $\eta$ directly corresponds to the power consumption of an ideal radio. Non-ideal radios are discussed in Section~\ref{app:radio_overheads}.

	It follows from the above that the duty-cycle of a tuple of sequences $B_\infty$, $C_\infty$, that are concatenations of finite length sequences $B$ and $C$ respectively, can be computed as follows.
	\begin{equation}
	\label{eq:etaRegular}
	\begin{array}{lrr}
	 \beta = \frac{\sum_{i = 1}^{m_B} \omega_i}{T_B} = \frac{\sum_{i = 1}^{m_B} \omega_i}{\sum_{i = 1}^{m_B} \lambda_i}, & \gamma = \frac{\sum_{i = 1}^{n_C} d_i}{T_C}, & \eta = \alpha \beta + \gamma
	\end{array}
	\end{equation}

\subsection{Beacon Length}
\label{sec:packet_Length}
\begin{figure}[tb]
	\centering
	\includegraphics[width=\imgWidth]{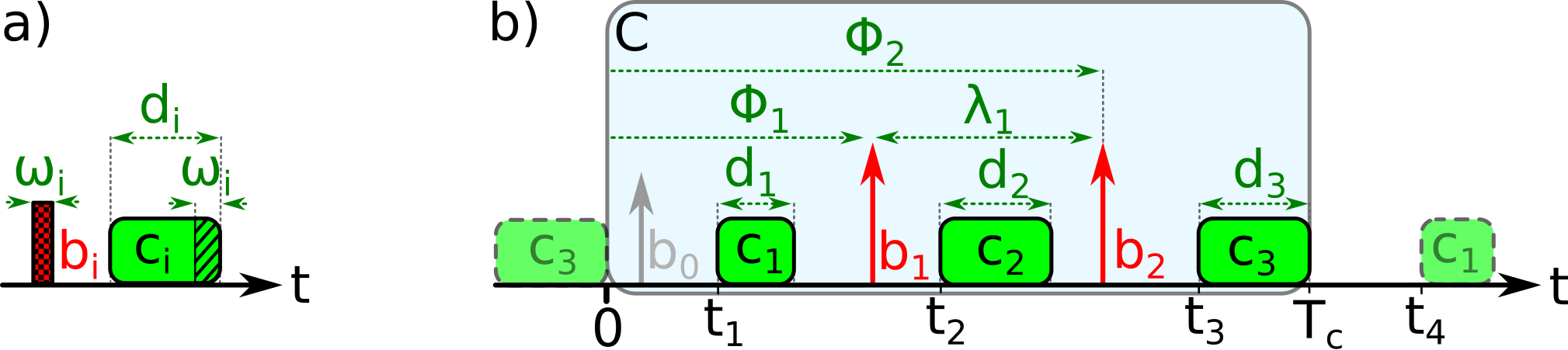}
	\caption{a) Beacons starting within the hatched area of window $c_i$ can only fractionally coincide. b) Offset $\Phi_1$ of the first beacon $b_1$ and $\Phi_2$ of the second beacon $b_2$ in range.}
	\label{fig:packetOverlap} 
\end{figure}
A beacon needs to be transmitted entirely within a reception window of a receiving device for being received successfully. Each beacon has a certain transmission duration $\omega_i$, and if the beacon transmission starts after the last $\omega_i$ time-units of a reception window (cf. after the start of the hatched area in Figure~\ref{fig:packetOverlap}a)), it cannot be received successfully. 
Nevertheless, for simplicity of exposition, for now we assume that any overlap between a beacon and a reception window leads to a successful discovery. We further assume that all beacons have the same length $\omega$ and neglect the contribution of the transmission duration of the first successfully received beacon to the worst-case latency. We study the relaxation of these assumptions in Section~\ref{app:non_zero_length_packets} and \ref{app:negelectLastBeacon}.

\section{Deterministic Beacon Sequences}
\label{sec:analysis}
A device \textit{F} can successfully discover another device \textit{E} only if \textit{E} sends a beacon during one of the reception windows of \textit{F}. We refer to the other direction as \textit{E} discovering \textit{F}. In what follows, we first consider \textit{F} discovering \textit{E} only, and later generalize it towards mutual discovery.

On device \textit{E}, let $B^\prime = b_1, b_2, ...$ be a subsequence of $B_\infty$. From here on, we will always assume that $b_1$ is the first beacon that is in range of a remote device \textit{F}. This is because any prior beacons of $B_\infty$, when \textit{E} is not within the range of \textit{F}, are not relevant for \ac{ND}.
Further, let \textit{F} run an infinite reception window sequence $C_{\infty}$. 
Though $B_\infty$ and hence $B^\prime \in B_\infty$ could be of infinite length, let us think of $B^\prime$ as a fixed-length sequence. This assumption is valid because in the case of a successful discovery, beacons that are sent thereafter are no longer relevant for the discovery procedure. 
Now recall that the reception windows of $C_\infty$ are formed by concatenations of a finite sequence $C$ and every instance of $C$ has its own time origin, as defined by Definition~\ref{def:chunk} (cf. Figure~\ref{fig:nd_protocol_rcv}). The first beacon $b_1$ in $B^\prime$ lies within a certain instance of $C$ and has a certain (random) offset $\Phi_1$ from the time origin of this instance of $C$. This is depicted in Figure~\ref{fig:packetOverlap}b), which shows an infinite beacon sequence consisting of concatenations of $C = c_1, c_2, c_3$, of which one full instance is depicted. In addition, the figure contains the last reception window $c_3$ of the preceding instance and the first reception window $c_1$ of the succeeding one. Further, three beacons $b_0$, $b_1$ and $b_2$ are shown, of which only $b_1$ and $b_2$ are in range. Here, $B^\prime$ consists of $b_1$, $b_2$ and some later beacons that are not shown in the figure. Beacon $b_1$ falls into the depicted instance of $C$ and has an offset of $\Phi_1$ time-units from its origin.

For some valuations of $\Phi_1$, at least one beacon of $B^\prime$ will coincide with a reception window of $C_\infty$. For other valuations of $\Phi_1$, there might be no beacon in $B^\prime$ that coincides with any reception window of $C_\infty$, irrespective of the length of $B^\prime$. If an overlapping pair of a beacon and a reception window exists for \textit{\textbf{all}} possible offsets $\Phi_1$, the tuple $(B^\prime,C_\infty)$ guarantees discovery within a bounded amount of time and hence realizes deterministic \ac{ND}. We, in the following, formalize the properties that such a tuple  $(B^\prime,C_\infty)$ needs to fulfill for guaranteeing discovery.


\subsection{Coverage and Determinism}
\label{sec:coverage_determinism}

A tuple ($C_\infty$, $B^\prime$), along with $\Phi_1$, is depicted in Figure~\ref{fig:packetOverlap}b).
For a given ($C_\infty$, $B^\prime$), it is obvious that the offset $\Phi_1$, which is a measure of the shift between $B^\prime$ and $C_\infty$, solely determines whether a beacon in $B^\prime$ overlaps with a reception window in $C_\infty$ or not. The time-duration after which such an overlap takes place, and hence the discovery latency, is also determined by $\Phi_1$.
For which values of $\Phi_1$ will beacon $b_1$ fall into one of the reception windows? Clearly, these are given by the set $\Omega_1 = \{[t_1, t_1 + d_1], [t_2, t_2 + d_2],...\}$ (cf. Figure~\ref{fig:packetOverlap}b)). In other words, if $\Phi_1$ lies within any interval belonging to $\Omega_1$, then $b_1$ is successfully received. 
Similarly, if $\Phi_2$ is the offset of $b_2$, then for $\Phi_2$ belonging to any interval in $\Omega_1$, $b_2$ will be successfully received (see Figure~\ref{fig:packetOverlap}b)).

Now, what are the offsets $\Phi_1$ of $b_1$, such that beacon $b_2$ is successfully received? These are given by the set  $\Omega_2 = \{[t_1 - \lambda_1, t_1 + d_1 - \lambda_1], [t_2 - \lambda_1, t_2 + d_2 - \lambda_1],..\}$, where $\lambda_1$ is the time-distance between the beacons $b_1$ and $b_2$, as already defined in Section~\ref{sec:nd_protocols} (see Figure~\ref{fig:packetOverlap}b)).
Therefore, $\Omega_2$ is obtained by shifting all elements of $\Omega_1$ by $\lambda_1$ time-units to the left. Similarly, $\Omega_3 = \{[t_1 - (\lambda_1+\lambda_2), t_1 + d_1 - (\lambda_1+\lambda_2)], [t_2 - (\lambda_1+\lambda_2), t_2 + d_2 - (\lambda_1+\lambda_2)],..\}$.
Then $\Omega_k$ for $k = 3,4,5,...$ is similarly defined as 
\begin{equation}
\scriptstyle\Omega_k = \{[t_1 - \sum_{i=1}^{k-1} \lambda_i, t_1 + d_1 - \sum_{i = 1}^{k-1} \lambda_i], [t_2 - \sum_{i=1}^{k-1} \lambda_i, t_2 + d_2 - \sum_{i=1}^{k-1} \lambda_i],...\}.\\
\end{equation}
Now consider a beacon sequence $B^\prime = b_1,...,b_m$ of length $m$. If $\Phi_1$ belongs to any interval in $\Omega_1 \cup \Omega_2 \cup ... \cup \Omega_m$, then one beacon from $B^\prime$ will be successfully received. 
We now extend this result and define a \textit{coverage map}, which can be used to reason about valuations of the initial beacon offset $\Phi_1$ that lead to successful discovery.
\nomenclature{$\Phi_{i}$}{Offset of the $i$'th beacon of a beacon sequence from the coordinate offset in $C$}
\nomenclature{$\Omega_i$}{Set of offsets $\Phi_1$ covered by beacon $i$}

\subsubsection{Coverage Maps}
\begin{figure}[bt]
	\centering
\begin{subfigure}{\imgWidth}
	\centering
	\includegraphics[width=\columnwidth]{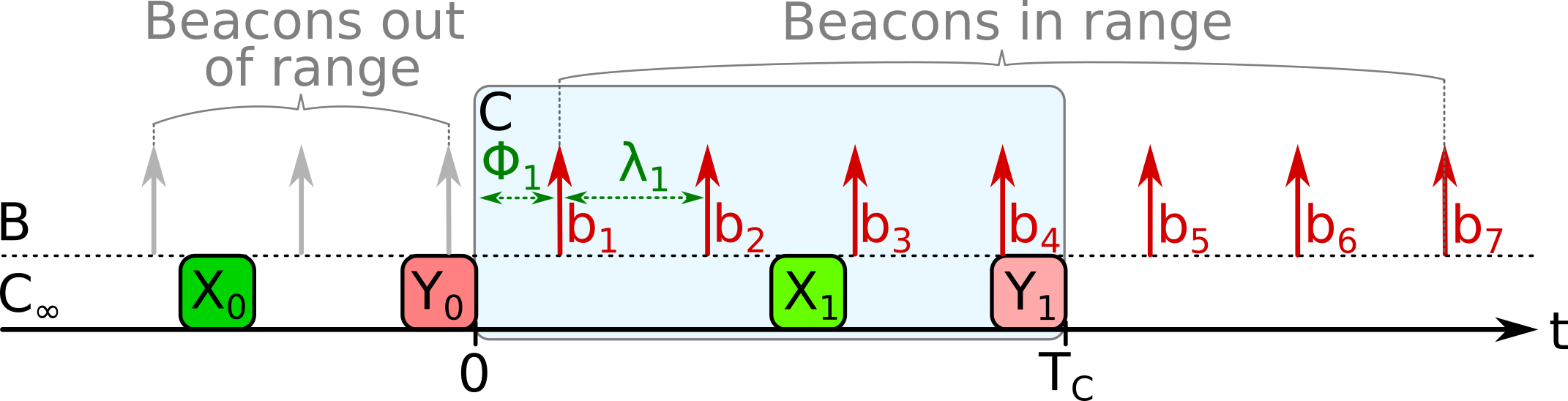}
	\caption{Example sequences $B^\prime = (b_1,...,b_7)$ and $C_\infty = (X_0, Y_0, X_1, Y_1,...)$.}
	\label{fig:coverage_maps_example}
 \end{subfigure}
\begin{subfigure}{\imgWidth}
	\centering
	\includegraphics[width=\columnwidth]{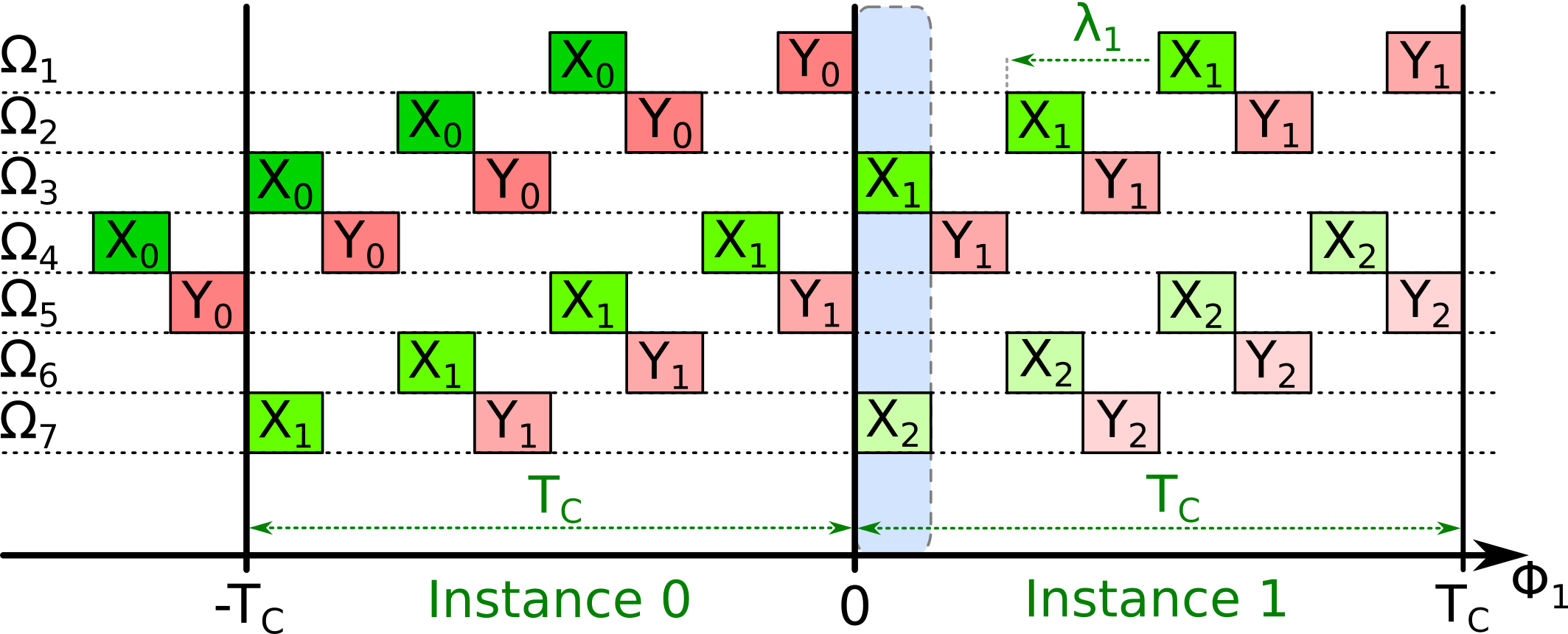}
	\caption{Coverage map for these sequences drawn over two periods.}
	\label{fig:coverage_maps} 
\end{subfigure}
\caption{Coverage maps.}
\label{fig:coverage_maps_whole}
\end{figure}
A coverage map is a formal representation of all offsets $\Phi_1$ for which any beacon in $B^\prime$ overlaps with a reception window in $C_\infty$. It also allows for a graphical representation, from which several properties of the tuple $(B^\prime,C_\infty)$ can be easily understood.

Recall that $C_\infty$ is a repeated concatenation of a sequence of reception windows $C$ (i.e., $C_\infty = C\mbox{ }C\mbox{ }C...$). Now, we need to be able to specify specific instances of $C$ within $C_\infty$. For this purpose, let us consider a simple example where $C$ has two reception windows $X$ and $Y$, and $C_\infty$ is therefore given by $C_\infty = X\mbox{ }Y\mbox{ }X\mbox{ }Y...$, and in order to distinguish between different instances of these reception windows, we will denote $C_\infty = X_0\mbox{ }Y_0\mbox{ }X_1\mbox{ }Y_1\mbox{ }X_2\mbox{ }Y_2...$ .
The reception windows $X_i$ and $X_{i+1}$, as well as $Y_i$ and $Y_{i+1}$, are $T_c$ time-units apart (see Figure~\ref{fig:coverage_maps_example} and also Figure~\ref{fig:nd_protocol_rcv}).

Figure~\ref{fig:coverage_maps_example} shows a sequence of beacons $B^\prime = b_1,...b_7$ from a transmitting device. Below, two reception windows $X_0, Y_0$ from a receiving device are depicted, together with their periodic repetitions $X_1$, $Y_1$, which are $T_C$ time-units later. 
Again, $b_1 \in B^\prime$ has a certain random offset $\Phi_1$ from the origin of $C$.
Figure~\ref{fig:coverage_maps} shows the coverage map for the sequences in Figure~\ref{fig:coverage_maps_example}. 

\begin{definition}[Covered]
An offset $\Phi_1$ is \textit{covered}, if at least one beacon in $B^\prime$ overlaps with any reception window in $C_\infty$ for this offset.
\end{definition}

Given the parameters of $(B^\prime, C_\infty)$, the \textit{construction} of a coverage map as in Figure~\ref{fig:coverage_maps}, is straightforward.
We believe that the notion of such a coverage map and its use go beyond deriving latency bounds as done in this paper. It would also be useful for analyzing and optimizing various kinds of different \ac{ND} protocols, including already known ones.


From coverage maps, we can derive the following properties.
\begin{asparaitem}
	\item \textbf{Beacon-to-beacon discovery latency $\mathbf{l^*}$}: For a given offset $\Phi_1$, let $l^*(\Phi_1)$ be the latency measured from the transmission time of the first beacon that is in range, to the first time a beacon is successfully received. 
	In Figure~\ref{fig:coverage_maps_whole}, $l^*(\Phi_1) = \tau_i - \tau_1 = \sum_{k=1}^{i-1} \lambda_{k}$, where $i$ is the smallest row number in which $\Phi_1$ is \textit{covered}.
	For example, for an offset $\Phi_1$ slightly above $0$ (i.e., an offset within the highlighted frame in Figure~\ref{fig:coverage_maps}), the beacon-to-beacon discovery latency will be $l^* = \tau_3 - \tau_1$, since $b_3$ is the earliest successful beacon for this offset.
  \item \textbf{Determinism}: By ensuring that all possible initial offsets are covered by at least one beacon, we can guarantee that $B^\prime$ is \textit{deterministic} with respect to $C_\infty$ (see next section for a formal definition of determinism).
  \item \textbf{Redundancy}: 
  For certain valuations of $\Phi_1$, one can see in Figure~\ref{fig:coverage_maps} that a beacon will be received by multiple reception windows. For example, for values of $\Phi_1$ within the shaded frame, beacons $b_3$ and $b_7$ will be received by the windows $X_1$ and $X_2$, respectively. By integrating over the length of all reception windows, for which such duplicate receptions happen, we can quantify the degree of redundancy of a tuple $(B^\prime, C_\infty)$.
  \end{asparaitem}
\nomenclature{$l^*$}{Beacon-to-beacon latency: Worst-case latency measured from the first beacon in range to the last, successfully received one.}
\subsubsection{Determinism}
Recall that protocols that can guarantee discovery for every possible initial offset are called \textit{deterministic}. This is formalized below. In particular, we distinguish between a \textit{beacon sequence} $B^\prime$ and a \textit{protocol} $(B_\infty, C_\infty)$ that can result in such a sequence.

\begin{definition}[Deterministic \ac{ND} Protocol]
\label{def:deterministic_nd_protocol}
A beacon sequence $B^\prime$ is \textit{deterministic} in conjunction with an infinite reception window sequence $C_\infty$, if all possible initial offsets $\Phi_1$ are covered by the tuple $(B^\prime, C_\infty)$. 
A \ac{ND} protocol $(B_\infty, C_\infty)$ is \textit{deterministic}, if for all $i$, $B^\prime_i = b_i, b_{i+1}, b_{i+2},...$ is a deterministic beacon sequence.
\end{definition}
Hence, deterministic \ac{ND} protocols $(B_\infty, C_\infty)$ always guarantee a bounded discovery latency, no matter when a beacon of $B_\infty$ comes within the range of a receiving device.
\begin{lemma}
	\label{lemma:periodiciyOfCoverageMaps}
	If a beacon sequence $B^\prime$ covers all offsets $\Phi_1$ within $[0, T_C]$, then all possible valuations of $\Phi_1$ are covered.
\pagebreak
	\begin{proof}
		Let us assume that a certain range of offsets \linebreak
		$[\Phi_x,\Phi_y]$, where $\Phi_x,\Phi_y \leq T_C$,  is covered by a beacon $b_i$ in conjunction with a certain reception window $c_j$. 
		Since the pattern of reception windows repeats every $T_C$ time-units, any $\Phi_1 \in [\Phi_x + T_C,\Phi_y + T_C]$ will result in $b_i$ being received by the reception window $c_{j + n_C}$, which is $T_C$ time-units after $c_j$.\qedhere
	\end{proof}
\end{lemma}

\begin{definition}[Redundant Sequences]
	\label{def:redundant}
	If any offset $\Phi_1$ within $[0,T_C]$ is covered by more than one beacon, then the tuple $(B^\prime,C_\infty)$ is \textit{redundant}. Otherwise, $(B^\prime,C_\infty)$ is \textit{disjoint}, since no intervals in the corresponding coverage map overlap.
\end{definition}
For example, in Figure~\ref{fig:coverage_maps}, all offsets $\Phi_1$ are covered and hence the corresponding tuple $(B^\prime,C_\infty)$ is deterministic. Further, since some offsets, e.g., the ones slightly above offset $0$ (marked by the highlighted frame in Figure~\ref{fig:coverage_maps}) are covered twice, it is also redundant.

\subsubsection{Coverage}
For a tuple $(B^\prime,C_\infty)$, certain values of $\Phi_1$ might be covered by multiple beacons, other values by exactly one beacon and yet others by no beacons. The notion of \textit{coverage} quantifies how different values of $\Phi_1 \in [0,T_C]$ are covered. To understand this, recall that $\Omega_i$ is a set of intervals. 
Let us now consider those (full or partial) intervals of $\Omega_i$ that lie within $[0, T_C]$. The sum of the lengths of all such intervals for all $\Omega_i$ captures a notion of \textit{coverage} that we formalize below.

\begin{definition}[Coverage]
\label{def:coverage}
Given a tuple $(B^\prime, C_\infty)$, let a certain offset $\Phi_1 \in [0,T_C]$ be covered by $k$ beacons, where $k \in \{0,1,2,...\}$. Let us define an auxiliary function $\Lambda^*(\Phi_1) = k$. Then, the \textit{coverage} $\Lambda$ is defined as
\begin{equation}
\Lambda = \int_{0}^{T_C} \Lambda^*(\Phi_1) d\Phi_1.
\end{equation}
\end{definition}
\nomenclature{$\Lambda^*(\Phi_1)$}{Number of beacons that cover the offset $\Phi_1$}
\nomenclature{$\Lambda$}{Coverage of a beacon sequence $B^\prime$ given an infinite reception window sequence $C_\infty$}
For example, in Figure~\ref{fig:coverage_maps}, if the lengths of $X_i$ and $Y_i$ are equal to unity and therefore $T_C = 8$, then $\Lambda = 14$.
If $\Lambda < T_C$, a tuple $(B^\prime, C_\infty)$ cannot be deterministic, which implies that for certain values of $\Phi_1$, no bounded discovery latency can be guaranteed. If $\Lambda = T_C$, then $(B^\prime, C_\infty)$ can either be deterministic and disjoint, or else, it will be redundant and not deterministic. If $\Lambda > T_C$, than $(B^\prime, C_\infty)$ cannot be disjoint, and may or may not be deterministic.

\subsection{Minimum Coverage}
While $\Lambda$ quantifies the coverage due to all beacons in $B^\prime$, we now quantify the coverage induced by individual beacons.
\begin{theorem}[Coverage per Beacon]
\label{thm:coverage_per_beacon}
Given a tuple $(B^\prime, C_\infty)$, every beacon $b_i \in B^\prime$ induces a coverage of exactly $\sum_{k=1}^{n_C} d_{k}$ time-units.
\begin{proof}
	The first beacon $b_1$ in $B^\prime$ will cover exactly those time-units for which $b_1$ directly coincides with a reception window. The sum of such matching offsets is therefore $\sum_{k=1}^{n_C} d_{k}$ time-units. Every later beacon $b_i$ will cover the same offsets shifted by the sum of beacon gaps $\sum_{k=1}^{i} \lambda_k$ to the left, which does not impact the amount of offsets covered. Since $C_\infty$ is an infinite concatenation of a finite sequence $C$, for every covered offset that is shifted out of the considered range $[0, T_C]$, the same amount from a later period is shifted into that range, such that each beacon $b_i$ covers exactly $\sum_{k=1}^{n_C} d_{k}$ time-units within $[0, T_C]$. \qedhere
\end{proof}
\end{theorem}
From the above, we are able to derive a minimum length of $B^\prime$.

\begin{theorem}[Beaconing Theorem]
	\label{thm:1stBeaconingTheorem}
	Given a tuple $(B^\prime, C_\infty)$, the minimum number of beacons $M$ a beacon sequence $B^\prime$ needs to consist of to guarantee deterministic discovery is:
	\begin{equation}
	\label{eq:NbMin}
	 M = \left \lceil \frac{T_C}{\sum_{k=1}^{n_C} d_{k}}\right\rceil
	 \end{equation}
	 \begin{proof}
	 	From Theorem~\ref{thm:coverage_per_beacon} it follows that every beacon induces a coverage of $\Lambda = \sum_{k=1}^{n_C} d_{k}$. For deterministic discovery, the coverage $\Lambda$ has to be at least $T_C$. Therefore, the number of beacons needed for deterministic \ac{ND} must be at least $\lceil \nicefrac{T_C}{\Lambda}\rceil$.\qedhere 
 	\end{proof}
\end{theorem}
\nomenclature{$M$}{Minimum number of beacons needed to deterministically match an infinite reception window sequence $C_\infty$}

It is worth mentioning that Theorem~\ref{thm:1stBeaconingTheorem} is a necessary, but not sufficient condition for deterministic \ac{ND}.
The positioning of the beacons, along with their number, together determine whether or not a tuple $(B^\prime, C_\infty)$ is deterministic.

\section{Fundamental Bounds}
\label{sec:bounds}
In this section, we derive the lower bounds on the worst-case latency that a \ac{ND} protocol could guarantee in different scenarios (e.g., symmetric or asymmetric discovery). In other words, given constraints like the duty-cycle, such a bound defines the best worst-case latency that any protocol could possibly realize.
First, we consider the most simple case in which one device \textit{F} runs an infinite reception window sequence $C_{F,\infty}$ without beaconing, whereas another device \textit{E} only runs an infinite beacon sequence $B_{E,\infty}$ without ever listening to the channel. We refer to this as \textit{unidirectional} beaconing.
\subsection{Bound on Unidirectional Beaconing}
\label{sec:uniDirBeaconing}
\begin{figure}[tb]
	\centering
	\includegraphics[width=\imgWidth]{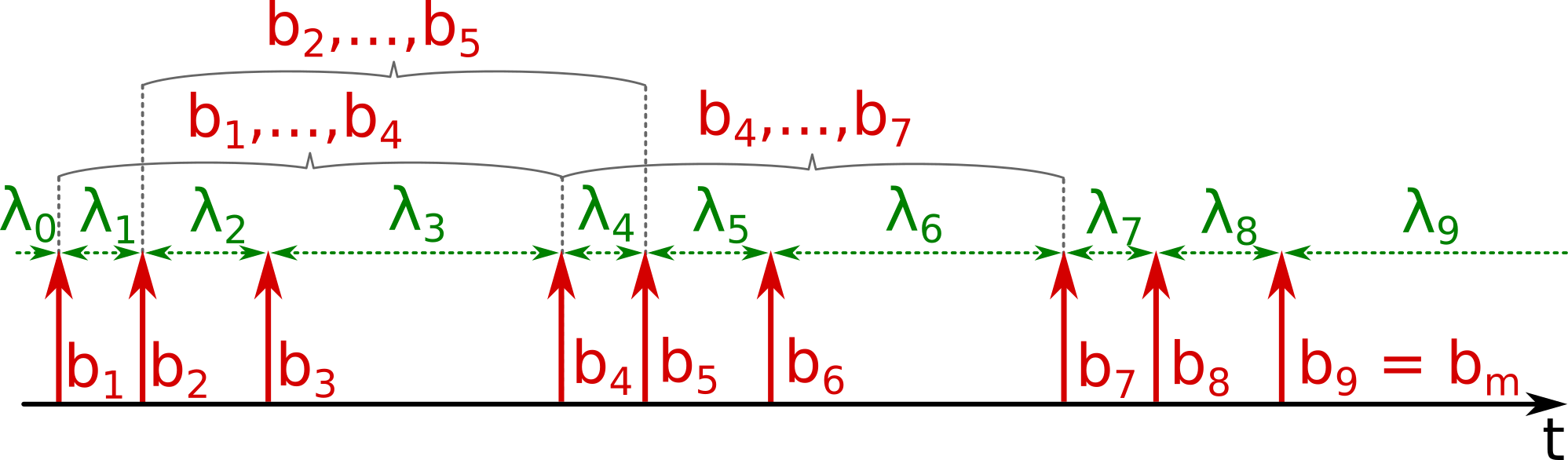}
	\caption{Partial sequences of an infinite beacon sequence.}
	\label{fig:proof_dm_r0} 
\end{figure}

\subsubsection{The Coverage Bound}
Consider a tuple $(B^\prime, C_\infty)$, where $B^\prime$ consists of $M$ beacons and $M$ is given by Theorem~\ref{thm:1stBeaconingTheorem}. 
Recall Theorem~\ref{thm:1stBeaconingTheorem} and the subsequent discussion. If $B^\prime$ is disjoint and deterministic, then for every value of $\Phi_1$, there is exactly one beacon in $B^\prime$ that overlaps with a reception window in $C_\infty$. What are the beacon gaps $\lambda_i$ using which such $M$ beacons need to be spaced for minimizing the discovery latency?

The worst-case beacon-to-beacon discovery latency $l^*$, measured from the first beacon in range to the earliest successfully received one, is given by the sum of the $M-1$ beacon gaps between these beacons. 
The first beacon in $B^\prime$ is the first beacon that was sent when the transmitter came within the range of the receiver. To measure the worst-case discovery latency $L$, time begins when the two devices come in range, which might be earlier than the time the first beacon in $B^\prime$ was sent.
How much earlier? At most by the beacon gap that precedes $B^\prime$. Recall that $B^\prime$ belongs to an infinite sequence $B_\infty$.
Hence, the lowest worst-case latency is achieved if the sum of these $M$ beacon gaps is minimized.
At the same time, all offsets in $[0, T_C]$ need to be covered exactly once for ensuring determinism. 

However, the following arguments rule out such $M$ consecutive beacon gaps to be arbitrarily short. $B_\infty$ has a transmission duty-cycle $\beta$, defined by the energy budget of the transmitter. Obviously, $\beta$ determines the average beacon gap $\overline{\lambda}$.
If the sum of a certain $M$ consecutive beacon gaps becomes smaller than $M \cdot \overline{\lambda}$, then the sum of a different $M$ consecutive beacon gaps within $B_\infty$ needs to exceed $M\cdot \overline{\lambda}$ in order to respect average beacon gap of $\overline{\lambda}$ defined by $\beta$.
Since any beacon in $B_\infty$ could be the first beacon in range, 
the $M$ beacons with the largest sum of beacon gaps determine the worst-case latency $L$.
Hence, in an optimal $B_\infty$, every sum of $M$ consecutive beacon gaps must be equal to $M \cdot \overline{\lambda}$.
It is worth noting that this requirement does not necessarily require equal beacon gaps, because the above property has to hold for a specific value of $M$ given by Theorem~\ref{thm:1stBeaconingTheorem}.
This is formalized in Lemma~\ref{lem:optimalRegularSchedules}. 

To illustrate the above, consider the following example.
Figure~\ref{fig:proof_dm_r0} shows a sequence $B^\prime = b_1,...,b_7$.
Here, let the minimum number $M$ of beacons for deterministic \ac{ND} be equal to $4$ and let the partial sequences $(b_1,...,b_4)$, $(b_2,...,b_5)$, $(b_4,...,b_7)$ be deterministic. 
Consider the sequence $b_1,...,b_4$. Let us assume that $b_4$ would be sent somewhat earlier than depicted. Then, by decreasing $\lambda_3$, the beacon gap $\lambda_4$ would increase accordingly, and though the sequence $b_1,...,b_4$ would result in a shorter discovery latency for all possible offsets, the sequence $b_4,...,b_7$ would lead to a larger worst-case latency.
The above observations are formalized below.
\begin{theorem}[Coverage Bound]
	\label{thm:worst_case_latency_theorem}
	The lowest worst-case latency that can be guaranteed by a tuple $(B_\infty, C_\infty)$ is:
	\begin{equation}
	\label{eq:wc_latency_theorem}
	L = \left \lceil \frac{T_C}{\sum_{i=1}^{n_C} d_{i}} \right \rceil \frac{\omega}{\beta}
	\end{equation}
\begin{prf}
		Consider a sequence $B^\prime = b_1,...,b_m$ with $m >> M$. 
		In $B^\prime$, if any sum of $M$ consecutive beacon gaps is less than $M \cdot \overline{\lambda}$, then the sum of a different $M$ consecutive beacon gaps will exceed  $M \cdot \overline{\lambda}$ and will define $L$. Since this is true for every $m$, it also holds for $B_\infty$.
	    The mean beacon gap is given by $\overline{\lambda} = \nicefrac{(\tau_m - \tau_1)}{(m-1)}$  and the worst-case latency by $L = M \cdot \overline{\lambda}$.
		Expressing the mean beacon gap by the duty-cycle for transmission (cf. Equation \ref{eq:duty_cycle_def}) and expanding $M$ using Theorem~\ref{thm:1stBeaconingTheorem} leads to Equation~\ref{eq:wc_latency_theorem}. \qedhere
	\end{prf}
\end{theorem}		 
	\begin{lemma}[Repetitive Beacon Sequences]
		\label{lem:optimalRegularSchedules}
		Given a repetitive $C_\infty$, every $B_\infty$ that guarantees the lowest worst-case latency is repetitive, with a period of $m_B = M$ beacons or $T_B = M \cdot \frac{\omega}{\beta}$ time-units.
	\end{lemma}

\subsubsection{Optimal Reception Window Sequences}
We know that in an optimal beacon sequence, the sum of every $M$ consecutive beacon gaps is $T_B$. The corresponding reception window sequence must be such that all offsets in $[0, T_C]$ are covered by such a beacon sequence. While there can be multiple such $C_\infty$ for a given $B_\infty$, the ones that are optimal must fulfill the following property.
\begin{theorem}[Overlap Theorem]
	\label{thm:TcOpt}
	For a given duty-cycle, every $C_\infty$ that minimizes the worst-case latency is given by the following equation.
	\begin{equation}
	\label{eq:TcOpt}
	T_C = k \cdot \sum_{i=1}^{n_C} d_i, \quad k \in \mathbb{N}
	\end{equation}
	\begin{prf}
		Let us assume that the length of $T_C$ is equal to $k \cdot \sum_{i=1}^{n_C} d_i - \Delta$, where $k$ is an integer and $\Delta \in [0,\sum_{i=1}^{n_C} d_i)$. Theorem~\ref{thm:worst_case_latency_theorem} implies the same worst-case latency for all values of $\Delta$, since the ceiling function in Equation~\ref{eq:wc_latency_theorem} does not change $L$. With $T_C = k \cdot \sum_{i=1}^{n_C} d_i - \Delta$, the reception duty-cycle is given by (cf. Equation \ref{eq:etaRegular}):
		\begin{equation}
			\label{eq:TcOptProof}
			\gamma = \frac{\sum_{i=1}^{n_C} d_i}{k \cdot \sum_{i=1}^{n_C} d_i - \Delta}
		\end{equation}
		From Equation \ref{eq:TcOptProof} follows that the duty-cycle is minimized when $\Delta = 0$, and hence $T_C = k \cdot\sum_{i=1}^{n_C} d_i$. \qedhere
	\end{prf}
\end{theorem}
The intuition behind Theorem~\ref{thm:TcOpt} is that if Equation~\ref{eq:TcOptProof} is not satisfied, then $T_C$ can be increased and therefore, the reception duty-cycle $\gamma$ can be reduced without requiring any additional beacons to guarantee discovery with the same $L$. 
In other words, the coverage intrinsically induced if Equation~\ref{eq:TcOptProof} is not satisfied exceeds what is needed for determinism.
By combining Theorem~\ref{thm:worst_case_latency_theorem} and \ref{thm:TcOpt}, we can derive a bound for unidirectional beaconing.

\begin{theorem}[Fundamental Bound for Unidirectional Beaconing]
	\label{thm:boundUnidir}
	Given a device \textit{E} that runs an infinite beacon sequence $B_{E,\infty}$ with a duty-cycle of $\beta_E$ and a device \textit{F} that runs an infinite reception window sequence $C_{F,\infty}$ with a duty-cycle of $\gamma_F$, the minimum worst-case latency that can be guaranteed for \textit{F} discovering \textit{E} is as follows.
		\begin{equation}
	\label{eq:boundUnidir}
	L = \left \lceil \frac{1}{\gamma_F} \right \rceil \frac{\omega}{\beta},
	\end{equation}
		Clearly, optimal values of $\gamma_F$ are of the form $\nicefrac{1}{k}$, $k \in \mathbb{N}$ and other values of $\gamma_F$ do not lead to an improved $L$ compared to them.
	
\begin{prf}
By combining $T_C = k \cdot \sum_{k=1}^n d_k$ from Theorem~\ref{thm:TcOpt} and Equation~\ref{eq:duty_cycle_def}, we can write Equation \ref{eq:wc_latency_theorem} as follows.
\begin{equation}
\label{eq:dmOptUnoptimized}
L = \frac{T_C}{\sum_{i=1}^{n_C} d_i} \cdot \frac{\omega}{\beta} = \frac{\omega}{\beta \cdot \gamma}
\end{equation}
This holds true for $\gamma_F$ in the form of $\nicefrac{1}{k}$, $k \in \mathbb{N}$. The proof for other duty-cycles follows from the above discussion.
\qedhere
\end{prf}
\end{theorem}

\subsection{Symmetric \ac{ND} Protocols}
In this section, we extend Theorem~\ref{thm:boundUnidir} towards bidirectional (i.e., device \textit{E} discovers device \textit{F} and vice-versa), symmetric (i.e., both devices \textit{E} and \textit{F} use the same duty-cycle $\eta$) \ac{ND}. For achieving bidirectional discovery, every device runs both a beacon and a reception window sequence, and we assume that $B_\infty$ and $C_\infty$ can be designed such that both sequences on the same device never overlap with each other. We relax this assumption in Appendix~\ref{app:same_sequences}.

\subsubsection{Bi-Directional Discovery}
We can achieve bidirectional discovery by running the optimal sequences $B_\infty$ and $C_\infty$ we have identified for unidirectional beaconing on both devices simultaneously. The latency of each partial discovery procedure (viz., the discovery of \textit{E} by \textit{F} and of \textit{F} by \textit{E}) is bounded by Theorem~\ref{thm:boundUnidir}. As a result, the worst-case latency for both partial discoveries being successful will also be bounded by Theorem~\ref{thm:boundUnidir}. Since both devices transmit and receive, we can optimize the share between $\beta$ and $\gamma$, which leads to the following bound.

\begin{theorem}[Symmetric Bound for Bi-Directional \ac{ND} Protocols]
	\label{thm:boundSymUncorrelated}
	For a given duty-cycle $\eta$, no bi-directional symmetric \ac{ND} protocol (i.e. every device runs the same tuple $(B_\infty, C_\infty)$) can guarantee a lower worst-case latency than the following.
	\begin{equation}
	\label{eq:boundSymUncorrelated}
	L = \min\Bigg(\underbrace{\left\lceil\frac{2}{\eta}\right\rceil^2 \cdot \frac{\omega \alpha}{\eta \left\lceil\frac{2}{\eta}\right\rceil - 1}}_{\mathbb{A}},\mbox{ }\underbrace{\left\lfloor\frac{2}{\eta}\right\rfloor^2 \cdot \frac{\omega \alpha}{\eta \left\lfloor\frac{2}{\eta}\right\rfloor - 1}}_{\mathbb{B}}\Bigg)
	\end{equation}
		
	\begin{prf}
	Because of Theorem~\ref{thm:TcOpt}, optimal reception duty-cycles are given by $\nicefrac{1}{\gamma} = k, k = 1,2,3,...$.
	By inserting $\eta = \alpha \beta + \gamma$ (cf. Definition~\ref{def:duty_cycle}) into Equation~\ref{eq:boundSymUncorrelated} and setting $\nicefrac{1}{\gamma} = k$, we obtain
	\begin{equation}
	\label{eq:L_by_k}
	L = \frac{k^2 \omega\alpha}{k \eta - 1}, k \in \mathbb{N}
	\end{equation}
	We now have to find the value of $k$ that minimizes $L$. 
	Let us for now allow non-integer values of $k$ in Equation~\ref{eq:L_by_k}. 
	By forming the first and second derivative of Equation~\ref{eq:L_by_k} by $k$, one can show that a local minimum of $L$ exists for $k = \nicefrac{2}{\eta}$, which is a non-integer number for most values of $\eta$. 
	By analyzing $\nicefrac{dL}{dk}$, we can further show that Equation~\ref{eq:L_by_k} is monotonically decreasing for values of $k < \nicefrac{2}{\eta}$ and monotonically increasing for values of $k > \nicefrac{2}{\eta}$. Hence, the only integer values of $k$ that potentially minimize $L$ are $\lceil \nicefrac{2}{\eta} \rceil$ and $\lfloor \nicefrac{2}{\eta} \rfloor$. 
	Inserting $k = \lceil \nicefrac{2}{\eta}\rceil$ or $k = \lfloor \nicefrac{2}{\eta}\rfloor$ into Equation~\ref{eq:dmOptUnoptimized} and taking the minimum latency among both possibilities leads to Equation~\ref{eq:boundSymUncorrelated}.
	\qedhere
	\end{prf}
\end{theorem}
In fact, Theorem \ref{thm:boundSymUncorrelated} also holds true for unidirectional beaconing, if the joint duty-cycle $\eta = \alpha \cdot \beta_E + \gamma_F$ of two devices \textit{E} and \textit{F} is to be optimized. Further, one can easily see that for small values of $\eta$, the ceiling function in Equation~\ref{eq:boundSymUncorrelated} only marginally affects the value of $L$, which can therefore be approximated by
\begin{equation}
\label{eq:eq:boundSymUncorrelatedSimplified}
L = \frac{4\alpha\omega}{\eta^2}.
\end{equation}
Even when both devices \textit{E} and \textit{F} transmit as well as receive, it is possible to design \textit{unidirectional} protocols in which only one of the two devices, \textit{E} or \textit{F}, can discover the other. Here, the beacons on both devices contribute to a joint notion of coverage, leading to a reduced latency bound compared to the case where both devices can discover each other mutually.
A bound for this possibility is given in Appendix~\ref{sec:correlatedSchedules}.

\subsubsection{Collision-Constrained Discovery}
\label{sec:collision_constrained_discovery}
For achieving the bound given by Theorem~\ref{thm:boundSymUncorrelated}, we have assumed that the beacons of multiple devices never collide. This assumption is reasonable for a pair of radios, in which collisions only rarely occur. However, as soon as more than two radios are carrying out the \ac{ND} procedure simultaneously, collisions become inevitable and some of the discovery attempts fail. As a result, some devices might discover each other after the theoretical worst-case latency has passed, or, depending on the protocol design, might not discover each other at all. Therefore, it is often required to limit the channel utilization and hence collision rate, which leads to an increased worst-case latency bound.

In protocols with disjoint sequences (i.e., every $\Phi_1$ is covered exactly once), every collision will lead to a failure of discovering within $L$. The collision probability is solely determined by the channel utilization $\beta$. We in this section study the worst-case latency that can be achieved if both $\eta$ and $\beta$ (and hence the collision probability) are given. We in addition discuss possibilities to reduce the number of failed discoveries for a given collision probability in Section~\ref{sec:multiDevDisc}.

Consider a number of $S$ senders, of which each occupies the channel by a time-fraction of $\beta$. The first beacon of an additional sender that starts transmitting (or comes into range) at any random point in time will face a collision probability of (cf. \cite{abramson:70}):
\begin{equation}
\label{eq:abramsonCollisions}
P_{c} = 1-e^{-2 (S - 1) \cdot \beta}
\end{equation}
\nomenclature{$S$}{Number of transmitting devices}
\nomenclature{$P_c$}{Collision probability}

Once a beacon has collided, the repetitiveness of infinite beacon sequences (cf. Lemma~\ref{lem:optimalRegularSchedules}) implies that the fraction of later beacons colliding with this device is predefined. 
Nevertheless, since all offsets between the two sequences occur with the same probability, the collision probability of every individual beacon is given by Equation~\ref{eq:abramsonCollisions}. 
When constraining the channel utilization to a maximum value $\beta_m$ that must never be exceeded, the following latency bound applies.
\begin{theorem}[Bound for Symmetric \ac{ND} with Constrained Channel Utilization]
\label{thm:boundSymChannelConstraint}
For a given upper bound on the channel utilization $\beta_m$, no symmetric \ac{ND} protocol can guarantee a lower worst-case latency than the following.
\begin{equation}
\label{eq:boundSymChannelConstraint}
L = \begin{cases}
\min(\mathbb{A},\mathbb{B}), & \text{if } \eta \leq \gamma_{o} + \alpha \beta_m\\[5pt]
\left \lceil \frac{1}{\eta  - \alpha \beta_m} \right \rceil \cdot \frac{\omega}{\beta_m},  & \text{if } \eta > \gamma_{o} + \alpha \beta_m
\end{cases} 
\end{equation}
Here, $\mathbb{A}$ and $\mathbb{B}$ are given by Equation~\ref{eq:boundSymUncorrelated} and $\gamma_{o} = \nicefrac{1}{\lceil \nicefrac{2}{\eta} \rceil}$, if $\mathbb{A} \leq \mathbb{B}$, and $ \nicefrac{1}{\lfloor \nicefrac{2}{\eta} \rfloor}$, otherwise.
\nomenclature{$\beta_m$}{Specified maximum channel utilization}
\begin{prf}
	 Given $\eta$, if the channel utilization that results from choosing the optimal value of $\gamma$ (see proof of Theorem~\ref{thm:boundSymUncorrelated}) does not exceed $\beta_m$, the bound given by Equation~\ref{eq:boundSymUncorrelated} remains unchanged. Otherwise, the bound is obtained from Equation \ref{eq:boundUnidir} by eliminating $\gamma$ using $\eta = \alpha \beta_m + \gamma$ (cf. Definition \ref{def:duty_cycle}).
\end{prf}
\end{theorem}

\subsection{Asymmetric Discovery}
So far, we have assumed that two devices \textit{E} and \textit{F} have the same duty-cycle, i.e., $\eta_E = \eta_F = \eta$. 
Next, we study the latencies of \textit{asymmetric} protocols with $\eta_E \ne \eta_F$. We thereby assume that each device knows the duty-cycle of and hence the sequences on its opposite device. This is relevant e.g., when connecting a gadget with limited power supply to a smartphone using BLE. Here, different sequences on both devices that account for the different power budgets can be determined in the specification documents of the service offered by the gadget. The case of every device being allowed to choose its duty-cycle autonomously during runtime is also relevant. The possible degradation of the optimal performance for this case needs to be studied in further work.
Whereas all our previously presented bounds are actually reachable by practical protocols, the bound we present for asymmetric \ac{ND} can only be reached for tuples of duty-cycles $(\eta_E, \eta_F)$, for which $\nicefrac{2}{\eta_F}$ and $\nicefrac{2}{\eta_E}$ are integers. For other duty-cycles, the achievable performance will lie slightly below. 

\begin{theorem}[Bound for Asymmetric \ac{ND}]
\label{thm:asymBound}
Consider two devices \textit{E} and \textit{F} with duty-cycles $\eta_E$ and $\eta_F$, where $\frac{2}{\eta_F}$ and $\frac{2}{\eta_E}$ are integers. The lowest worst-case latency for two-way discovery is as follows.
\begin{equation}
\label{eq:boundAsym}
L = \frac{4 \alpha \omega}{\eta_E \eta_F}
\end{equation}
\begin{prf}
According to Theorem~\ref{thm:boundUnidir},  if $\nicefrac{1}{\gamma_E}$ and $\nicefrac{1}{\gamma_F}$ are integers, the lowest worst-case one-way discovery latency $L_F$ for device \textit{F} discovering device \textit{E} and the latency $L_E$ for the reverse direction are as follows.
\begin{equation}
\label{eq:boundAsymSameWcBounds}
\begin{array}{lr}
L_F = \frac{\omega}{\gamma_F \cdot \beta_E}, & L_E = \frac{\omega}{\gamma_E \cdot \beta_F}
\end{array}
\end{equation}
The global worst-case latency for two-way discovery is given by $L = max(L_E, L_F)$.
Because of this, every optimal asymmetric \ac{ND} protocol must fulfill $L_F = L_E$, since in cases of e.g., $L_F > L_E$, one could decrease the reception duty-cycle $\gamma_F$ of device \textit{F} and still achieve the same two-way discovery latency $L$. 
From $L_E = L_F$ and Equation~\ref{eq:boundAsymSameWcBounds} follows that $\nicefrac{\gamma_F}{\gamma_E} = \nicefrac{\beta_F}{\beta_E}  = const = \mu$.
By substituting $\beta_E$ by $\nicefrac{\beta_F}{\mu}$ in $L_F$ (cf. Equation \ref{eq:boundAsymSameWcBounds}) and by substituting $\gamma_F = \eta_F - \alpha \beta_F$, we obtain:
\begin{equation}
L_F = \frac{\omega \mu}{(\eta_F - \alpha \beta_F)\beta_F}
\end{equation}
\nomenclature{$\mu$}{Constant ratio of the reception or beaconing duty-cycles of two devices}
By differentiating $L_F$ by $\beta_F$, we can show that $L_F$ is minimal for $\beta_F = \nicefrac{\eta_F}{2 \alpha}$ and hence $\gamma_F = \nicefrac{\eta_F}{2}$. Similarly, $L_E$  has a local minimum at $\beta_E = \nicefrac{\eta_E}{2 \alpha}$. 
We note that if $\nicefrac{2}{\eta_E}$ and $\nicefrac{2}{\eta_F}$ are integers, also $\nicefrac{1}{\gamma_E}$ and $\nicefrac{1}{\gamma_F} 	$ are integers. 
When re-substituting $\mu$ by $\nicefrac{\beta_F}{\beta_E}$ and replacing $\beta_F$ and $\beta_E$ by their optimal values, we obtain Equation~\ref{eq:boundAsymSameWcBounds}.
\end{prf}
\end{theorem}

\section{Relaxation of Assumptions}
\label{sec:validity_of_assumptions}
In Section~\ref{sec:analysis}, for the sake of ease of presentation, we have made multiple simplifying assumptions. In this Section, we relax all assumptions that have an impact on the discovery latency, study how the fundamental bounds are impacted by this and numerically evaluate the difference between the ideal and real bounds. In particular, we consider the bound for unidirectional beaconing from Theorem~\ref{thm:boundUnidir}. We thereby consider only optimal reception duty-cycles $\gamma$, since other values of $\gamma$ do not lead to any improvement of $L$ (see Theorem~\ref{thm:boundUnidir}).




\subsection{Successful Reception of All Beacons}
\label{app:non_zero_length_packets}

Throughout the paper, we have assumed that also beacons that only partially overlap with a reception window are received successfully.	
To account for the fact that beacons cannot be received if their transmissions start within the last $\omega$ time-units of each reception window (since they must entirely overlap with the window), we have to artificially shorten the actual length of each reception window $d_k$ by one beacon transmission duration $\omega$ when computing discovery latencies, while still accounting for the full length of each reception window in computations of the duty-cycle.  
As a result, the coverage per beacon $\Lambda$ in Equation \ref{eq:wc_latency_theorem} from Theorem \ref{thm:worst_case_latency_theorem} needs to be reduced by one beacon transmission duration $\omega$ for each reception window, such that a modified bound can be given as follows.
	\begin{equation}
\label{eq:wc_latency_theorem_nonZeroPackets}
L = \left \lceil \frac{T_C}{\sum_{k=1}^{n_C} (d_{k} - \omega)} \right \rceil \frac{\omega}{\beta} 
\end{equation}
Clearly, this increases the worst-case latency that can be achieved for a given reception duty-cycle $\gamma = \nicefrac{\sum_{k=1}^{n_C} d_k}{T_C}$. From this and Equation \ref{eq:wc_latency_theorem_nonZeroPackets} follows that for a given reception duty-cycle $\gamma$, the increase of $L$ becomes larger for higher numbers of reception windows $n_C$ per period $T_C$. Hence, the tightest bound can be achieved for $n_C = 1$, for which the term $\sum_{k=1}^{n_C} d_k$ becomes $d_1$. 
Using this and by restricting $\gamma$ to optimal values and hence setting $T_C = k \cdot (d_1 - \omega)$ (cf. Theorem~\ref{thm:TcOpt}), we can write Equation \ref{eq:wc_latency_theorem_nonZeroPackets} as:
\begin{equation}
\label{eq:wc_latency_theorem_nonZeroPackets_one_chunk}
L = \frac{T_C \omega}{T_C \beta \gamma - \beta \omega}
\end{equation}
By examining the first derivative of Equation \ref{eq:wc_latency_theorem_nonZeroPackets_one_chunk}, one can show that $L$ becomes smaller for growing values of $T_C$. However, $L$ cannot become arbitrarily large. If $n_C = 1$ (i.e., one window per reception period), $T_C$ cannot exceed $L$ time-units because of the following reason. Consider a beacon that is sent at the very beginning of a reception window of another device. Let us assume that two devices come into range infinitesimally after this beacon has been sent. Since optimal beacon sequences do not contain more than one beacon within $L$ time-units that overlap the same reception window, the next successful beacon will overlap with the subsequent reception window, which begins $T_C$ time-units later. In other words, $T_C$ must not exceed $L$ time-units, since $L$ would otherwise scale with $T_C$. For $n_C > 1$, it is $T_C \leq n_C L$. Hence, we can to substitute $T_C$ by $L$ in Equation~\ref{eq:wc_latency_theorem_nonZeroPackets_one_chunk}.
Solving this equation by $L$ leads to the following bound:
\begin{equation}
 L = \frac{\omega + \beta \omega}{\beta \gamma}
\end{equation}
One can show that this bound is independent of the number of reception windows $n_C$ per reception period. 

\subsection{Neglecting the First Successful Beacon}
\label{app:negelectLastBeacon}
Throughout the paper, we have neglected the transmission duration of the first successfully received beacon.
We can account for this by adding $\omega$ time-units to Equation \ref{eq:wc_latency_theorem_nonZeroPackets}.
By forming the first and second derivative, we can show that the optimal share between transmission and reception is not influenced by this. When accounting for this beacon, all our presented bounds become by $\omega$ time-units longer (e.g., Theorem \ref{thm:boundSymUncorrelated} becomes $L = \nicefrac{4 \alpha \omega}{\eta^2} + \omega$). Besides from this, there are no changes, since finding the optimal beaconing duty-cycle $\beta$ is the only step that is potentially sensitive on adding $\omega$ to $L$. 

\subsection{Radio Overheads}
\label{app:radio_overheads}
Throughout this paper, we have assumed that the radios do not require any energy to switch from sleep mode to transmission or reception, and vice-versa. We now assume an overhead $d_{oTx}$ to switch the radio from the sleep mode to transmission and back, and an overhead $d_{oRx}$ to switch from the sleep mode to reception and back. These overheads can be regarded as effective durations of additional active time, i.e., as the actual durations that are needed to switch the radio's mode of operation, weighted by the quotient of the average power consumption during the switching phase over the power consumption for reception. For the sake of simplicity of exposition, we also assume the same overheads for switching directly between reception and transmission, without going to a sleep mode in between.

When accounting for these offsets, one can derive from Equation~\ref{eq:etaRegular} that the duty-cycle for transmission $\beta$ and for reception $\gamma$ become: 
\begin{equation}
\begin{array}{lr}
\beta = \sum_{i=1}^{m_B} \frac{\omega_i + d_{oTx}}{\lambda_i}, & \gamma = \frac{\sum_{k=1}^{n_C}(d_k + d_{oRx})}{T_C}\\
\end{array}
\end{equation}
With this, Equation~\ref{eq:boundUnidir}, from which all other bounds are derived, becomes:
\begin{equation}
L = \left\lceil\frac{1}{\gamma - \frac{n_C d_{oRx}}{T_C}}\right\rceil \cdot \frac{\omega + d_{oTx}}{\beta}
\end{equation}
We only consider optimal values of $\gamma$, which are given by $\gamma = \nicefrac{1}{k} + \nicefrac{n_C d_{oRx}}{T_C}$.
Large values of $T_C$ minimize $L$, and hence, as explained in Section~\ref{app:non_zero_length_packets}, we set $T_C = n_C \cdot L$. This leads to the following bound.
\begin{equation}
L = \frac{d_{oTx} + \omega + \beta \cdot d_{oRx}}{\beta \gamma}
\end{equation}
\nomenclature{$d_{oTx}$}{Effective additional active time for switching from sleep to transmission and vice-versa}
\nomenclature{$d_{oRx}$}{Effective additional active time for switching from sleep to reception and vice-versa}

\subsection{Evaluation}
In this section, we numerically evaluate the impact of the simplifying assumptions described above on the latency bound.

We assume a transmission duration of $\SI{32}{\micro s}$, which corresponds to a 4-byte beacon on a $\SI{1}{MBit/s}$ - radio used for e.g, \ac{BLE}. We consider a range of duty-cycles $\beta$ of the sender and $\gamma$ of the receiver between $\SI{0.055}{\percent}$ and $\SI{5.55}{\percent}$.
This range of duty-cycles leads to a practically relevant range of discovery latencies from $\SI{0.1}{s}$ to $\SI{100}{s}$ for optimal protocols on ideal hardware platforms (i.e., for Equation~\ref{eq:boundUnidir}). We only consider optimal values of $\gamma$. We further assume $\alpha = 1$.

\begin{figure}[tbh]
	\centering
	\includegraphics[width=\imgWidth]{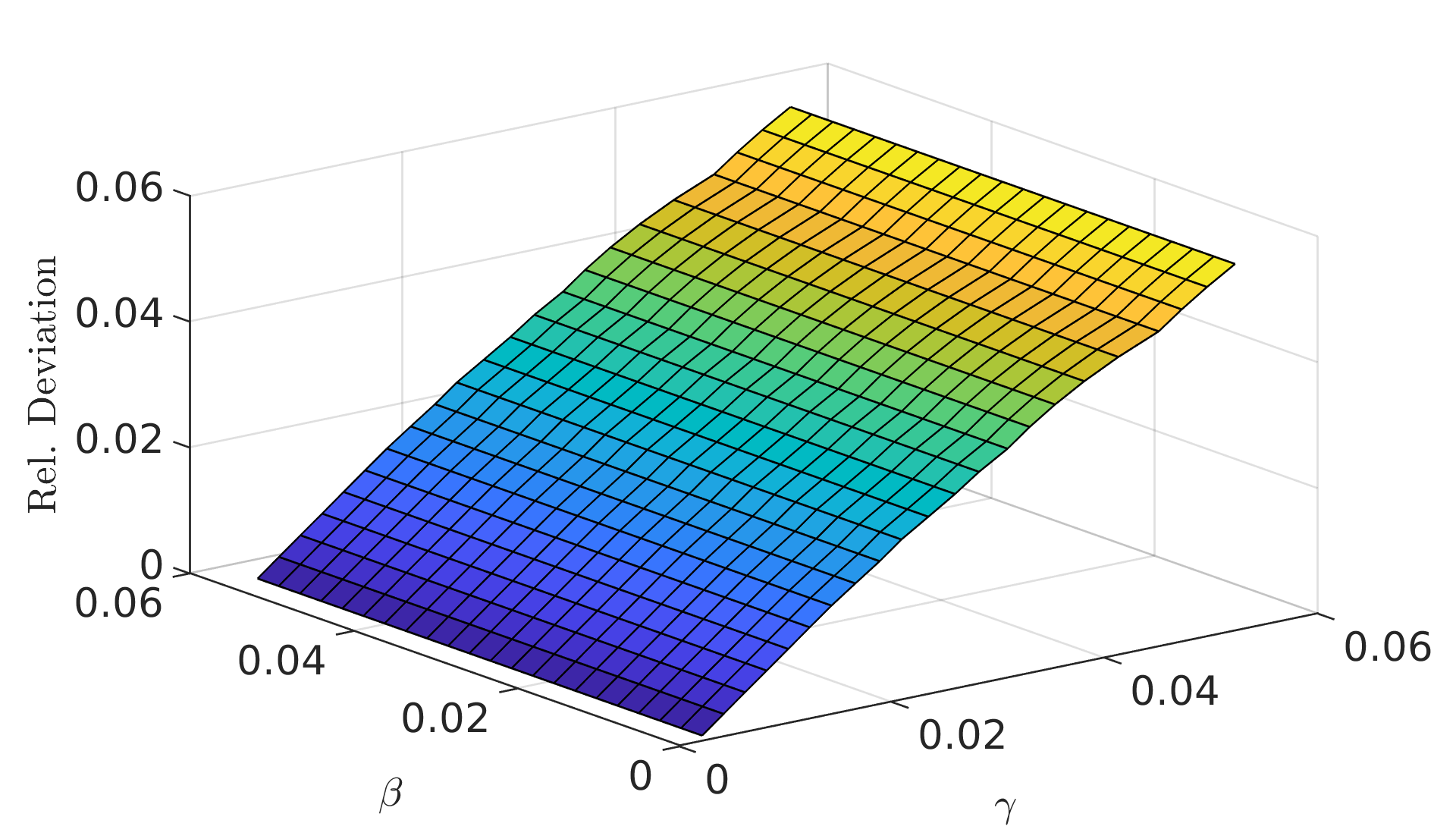}
	\caption{Relative difference between real and ideal bound on radios without switching overheads.}
	\label{fig:realBoundDiff_ideal}
 \end{figure}

Let $L_i$ denote the ideal latency bound (viz., Equation~\ref{eq:boundUnidir}) and $L_r$ the latency bound with relaxed assumptions. 
As can be seen from Figure~\ref{fig:realBoundDiff_ideal}, in the considered range of duty-cycles, the relative deviation $\nicefrac{(L_r - L_i)}{L_i}$ ranges between nearly $\SI{0}{\percent}$ to $\SI{6}{\percent}$.
\nomenclature{$L_i$}{Ideal worst-case discovery latency}
\nomenclature{$L_r$}{Worst-case discovery latency when relaxing all simplifying assumptions}

While Figure~\ref{fig:realBoundDiff_ideal} provides a platform-independent comparison for any ideal $\SI{1}{MBit/s}$ radio, what performance can be achieved on existing hardware platforms? For a Nordic nRF51822 SOC~\cite{nrf51822:14}, the switching overheads are approximately given by $d_{oRx} = d_{oTx} = \SI{140}{\micro s}$. Within the considered range of duty-cycles, the relative deviation to the ideal bound ranges between $\SI{438}{\percent}$ and $\SI{467}{\percent}$.
\section{Previously Known Protocols}
\label{sec:previously_known_bounds}

In this section, we relate the worst-case performance of popular protocols and previously known bounds to the fundamental limits described in the previous section. Due to their relevance in practice, we consider only small duty-cycles $\eta$, for which the bound for symmetric protocols is given by Equation~\ref{eq:eq:boundSymUncorrelatedSimplified}.

\subsection{Worst-Case Bound of Slotted Protocols}
\begin{figure}[tb]
	\centering
	\includegraphics[width=\imgWidth]{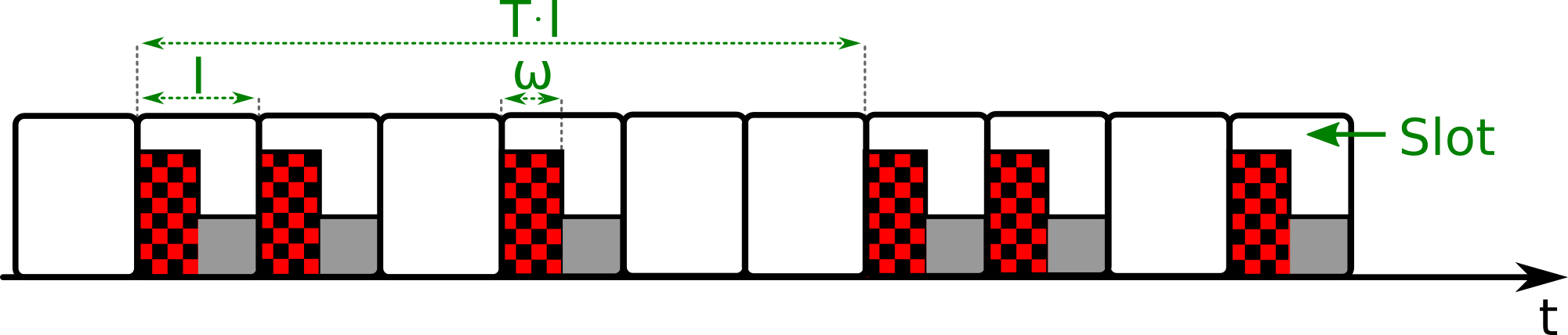}
	\caption{Slotted schedule proposed in \cite{zheng:06}. Hatched bars depict beacons, smaller rectangles reception windows.}
	\label{fig:zhengBoundSchedule} 
\end{figure}
As already described in Section \ref{sec:related_work}, a worst-case number of slots within which discovery can be guaranteed is known for slotted protocols \cite{zheng:03, zheng:06}. The corresponding worst-case latency in terms of time is proportional to the slot length $I$, for which there is no known lower limit.
In this section, we for the first time transform this worst-case number of slots into a latency bound and establish the relations to the fundamental bounds on \ac{ND} presented in this paper. We will also address the bound presented in \cite{meng:14, meng:16}, which has been claimed to be tighter than the bound in \cite{zheng:03, zheng:06}.

\subsubsection{Latency/Duty-Cycle Bound}
According to \cite{zheng:03, zheng:06}, no symmetric slotted protocol can guarantee discovery within $T$ slots by using less than $k \geq \sqrt{T}$ active slots per $T$. The associated worst-case latency $L$ is $T \cdot I$ time-units, which is directly proportional to the slot length $I$. We in the following derive a theoretical lower limit for $I$ and hence for $L$.

Slotted protocols can only function properly if the beacon length $\omega$ is``at least one order of magnitude smaller than $I$'' \cite{zheng:06}. If this requirement is not fulfilled, often a beacon might not overlap with a reception window even though the active slots of two devices overlap, as 
illustrated in Figure~\ref{fig:zhengBoundSchedule}. Here, the slot length $I$ in a slot design as proposed in \cite{zheng:06} has been set to $2 \cdot \omega$. As can be seen, practically none of the offsets for which two active slots overlap would lead to a successful reception, since every beacon would only partially overlap with a reception window. If $I$ would be increased, the fraction of successful offsets would gradually become larger. For achieving zero collisions independently of the slot length, let us assume a full duplex radio, which can both transmit and receive during the same points in time. 
Then, the theoretical limit on the slot length $I$ becomes as low as one beacon transmission duration $\omega$, which leads to the following duty-cycle:
\begin{equation}
\label{eq:dutyCycleSlottedFullDuplex}
\eta = \frac{k\cdot(I +  \alpha \omega)}{T \cdot I} = \frac{k\cdot(I +  \alpha \omega)}{L}
\end{equation}
Since the limit from \cite{zheng:03, zheng:06} requires that $k \geq \sqrt{T} = \sqrt{\nicefrac{L}{I}}$, with a slot length of $I = \omega$, Equation~\ref{eq:dutyCycleSlottedFullDuplex} leads to the following latency limit:
\begin{equation}
\label{eq:bundSlotted}
L \geq \frac{\omega(1 + 2 \alpha + \alpha^2)}{\eta^2}
\end{equation}
For $\alpha = 1$, this bound becomes $\frac{4 \omega}{\eta^2}$ and hence identical to the fundamental bound for symmetric protocols given by Theorem~\ref{thm:boundSymUncorrelated}. For all other values of $\alpha$, this bound exceeds the one given by Theorem~\ref{thm:boundSymUncorrelated}.

However, the assumption of full-duplex radios is not fulfilled by most wireless devices. Further, every wireless radio requires a turnaround time to switch from transmission to reception, during which the radio is unable to receive any beacons. 
Even for recent radios, this time is large against the beacon transmission duration $\omega$ (e.g, for the nRF51822 radio \cite{nrf51822:14}, it lies around $\SI{140}{\micro s}$, whereas beacons can be as short as $\SI{32}{\micro s}$). 
Therefore, $I$ will be orders of magnitude larger than $\omega$, which linearly increases the worst-case latency slotted protocols can guarantee in practice. It is worth mentioning that this increase occurs in addition to the duty-cycle overhead induced by the turnaround times of the radio.

We now study the bound presented in \cite{meng:14, meng:16}, which has been claimed to be lower in terms of slots than the one presented in \cite{zheng:03, zheng:06}. It is achieved by assuming two beacon transmissions per active slot (\cite{zheng:03, zheng:06} assumes only one), of which one beacon is sent slightly outside of the slot boundaries. 
By accounting for the two beacons per active slot, Equation~\ref{eq:dutyCycleSlottedFullDuplex} becomes $\eta = \frac{k\cdot(I +  2 \alpha \omega)}{L}$, which leads to the following bound for the protocols proposed in \cite{meng:14, meng:16}:
\begin{equation}
\label{eq:boundMengDiffCodes}
L \geq \frac{\omega(\frac{1}{2} + 2 \alpha + 2 \alpha^2)}{\eta^2} 
\end{equation}
This bound becomes minimal for $\alpha = \nicefrac{1}{2}$, for which it is identical to the bound in Theorem~\ref{thm:boundSymUncorrelated}. Hence, the bound in \cite{meng:14, meng:16} is lower in terms of slots than the bound in \cite{zheng:03, zheng:06}, but indentical or larger in terms of time. 

\subsubsection{Latency/Duty-Cycle/Channel Utilization Bound}
All previously known bounds for slotted protocols are in the form of relations between the worst-case number of slots and the duty-cycle. The channel utilization, which is directly related to the beacon collision rate, has not been considered before. However, in slotted protocols, the channel utilization depends both on the number of active slots per period and on the slot length. For sufficiently large slot lengths, the turnaround times of the radio only play a negligible role. Further, the time for reception in each slot approaches nearly the whole slot length $I$. Hence, for $I >> \omega$, we can compute the duty-cycle of slotted protocols as follows.
\begin{equation}
\label{eq:dutyCycleSlotted}
\beta = \frac{k \omega}{I T}, \quad \gamma = \frac{k I}{I T} = \frac{k}{T}, \quad \eta = \gamma + \alpha \beta\\
\end{equation}
With the requirement of $k \geq \sqrt{T}$ from \cite{zheng:03, zheng:06}, one can express the slot length $I$ by the desired channel utilization $\beta$ in Equation~\ref{eq:dutyCycleSlotted}, which results in the following bound.
\begin{equation}
\label{eq:boundSlottedChanUtil}
L \geq \frac{\omega}{\eta \beta - \alpha \beta^2}
\end{equation}
From comparing Theorem~\ref{thm:boundSymChannelConstraint} (cf. Equation~\ref{eq:boundSymChannelConstraint}) to Equation~\ref{eq:boundSlottedChanUtil}, it follows that if $\beta_m$ lies below $\nicefrac{\eta}{2 \alpha}$, the worst-case latency a slotted protocol can guarantee with a channel-utilization of $\beta = \beta_m$ is identical to the corresponding fundamental bound (recall that we only consider optimal duty-cycles). For $\beta_m > \nicefrac{\eta}{2 \alpha}$, slotted protocols cannot reach the fundamental bound from Theorem~\ref{thm:boundSymChannelConstraint}.
In practice, this means that slotted protocols can potentially perform optimally in busy networks with many devices discovering each other simultaneously, but cannot offer optimal performance in networks in which new devices join gradually and hence only a master node and the joining device need to carry out \ac{ND} at the same time.
\nomenclature{$I$}{Slot length in a slotted protocol}

We in the following evaluate the popular protocols Disco~\cite{dutta:08}, Searchlight-Striped~\cite{bakht:12}, U-Connect~\cite{Kandhalu:10} and diffcode-based protocols~\cite{zheng:03} and compare them to the fundamental bound given by Theorem~\ref{thm:boundSymChannelConstraint}. In Disco, active slots are repeated after every $p_1$ and $p_2$ slots, where $p_1$ and $p_2$ are coprimal numbers. The Chinese Remainder Theorem  implies that there is a pair of overlapping slots among two devices every $p_1 \cdot p_2$ time-units. U-Connect also relies on coprimal numbers for achieving determinism. In contrast, Seachlight defines a period of $T$ and a hyper-period of $T^2$ slots. The first slot of each period is active, whereas a second active slot per period systematically changes its position, until all possible positions have been probed. Diffcode-based solutions are built on the theory of block designs and hence guarantee a pair of overlapping slots among two devices with the minimum possible number of active slots per worst-case latency. More details on these protocols can be found in ~\cite{chen:16}.

Slot length-dependent equations on the worst-case latency and duty-cycle of these protocols are available from the literature. When assuming sufficiently large slots and by expressing the slot length $I$ by the channel utilization $\beta$ similarly to Equation~\ref{eq:dutyCycleSlotted}, one can derive the relations between the worst-case latency, duty-cycle and channel utilization given in Table~\ref{tab:slottedPerformance}. Clearly, only Diffcode-based schedules reach the optimal performance in this metric, whereas all other ones perform below the optimum.
 
In summary, slotted protocols can perform optimal in the laten-cy/duty-cycle/channel utilization metric, if the channel utilization remains low. In the latency/duty-cycle metric, however, higher required channel utilizations prevent slotted protocols from performing optimally.

\begin{table}
	\begin{center}
		{\renewcommand{\arraystretch}{1.3}
			\begin{tabular}{cc}
				\hline Protocol & $L(\beta, \eta)$ \\ 
				\hline 	Diffcodes \cite{zheng:03}& $\frac{\omega}{\eta \beta - \alpha \beta^2}$\\
				Disco \cite{dutta:08}& $\frac{8 \omega}{\eta \beta - \alpha \beta^2}$ \\ 
				Searchlight-S \cite{bakht:12}& $\frac{2 \omega}{\eta \beta - \alpha \beta^2}$\\
				U-Connect \cite{Kandhalu:10}& $\frac{\left(3 \omega + \sqrt{\omega^2 (8 \eta - 8 \alpha \beta + 9)}\right)^2}{8 \omega \beta \eta - 8 \omega \alpha \beta^2}$ \\ 
				\hline 
			\end{tabular} 
		}
	\end{center}
	\caption{Worst-case latencies of slotted protocols.}
	\label{tab:slottedPerformance}
\end{table}
\vspace*{2em}
\subsection{Worst-Case Bound of Slotless Protocols}
In slotted protocols, the number of beacons is always coupled to the number of reception phases. Slotless protocols are not subjected to this constraint. Can they reach optimal latency/duty-cycle relations?
In \cite{kindt:18}, two parametrization schemes for slotted protocols, called $PI-0M$ and $PI-kM^+$, have been proposed, which have been claimed to provide the best latency/duty-cycle performance among all known slotless protocols. We therefore in the following relate their performance to the bounds presented in Section~\ref{sec:bounds}.

In such slotless protocols, beacons are sent periodically with a period $T_B$ and the device listens to the channel for $d$ time-units once per period $T_C$. When optimizing $T_B$, $T_C$ and $d$, the worst-case latency is given by (cf. \cite{kindt:18} for details):
\begin{equation}
	L = \left ( \left \lceil \frac{T_C - d + \omega}{T_B} \right \rceil + 1\right)\cdot T_B + \omega
\end{equation}
As for our bounds, we assume that 1) beacons that are sent within the last $\omega$ time-units of each reception phase are received successfully and 2) the transmission duration of the first successfully received beacon is neglected. Under these assumptions, we can set $\omega = 0$ and obtain the following worst-case latency of $PI-0M$:
\begin{equation}
	\label{eq:equationPIkmRaw}
	L = \left ( \left \lceil \frac{T_C - d}{T_B} \right \rceil + 1\right)\cdot T_B
\end{equation}
Further, \cite{kindt:18} requires that $T_B = d$ and ${T_C = (M+1) d - \Delta}$, ${M \in \mathbb{R}}$ and $\Delta \to 0$. This leads to a worst-case latency $L$ of $\frac{\omega (M + 1)^2}{\eta (M+1)-1}$ time-units. By forming the first and second derivative of $L$, one can find the optimal value of $M$, using which $L$ becomes $\nicefrac{4 \omega \alpha}{\eta^2}$. This is identical to Theorem~\ref{thm:boundSymUncorrelated} and hence, under the assumptions described above, the $PI-0M$ scheme is optimal in the latency/duty-cycle metric. One can also show that under ideal assumptions, $PI-kM^+$ performs optimally, while it performs slightly below $PI-0M$ under relaxed assumptions.
Which degradation of the latency bound of the $PI-0M$ scheme do these assumptions imply in practice? When assuming a beacon transmission duration of $\omega = \SI{32}{\micro s}$ and a range of duty-cycles between $\SI{0.1}{\percent}$ and $\SI{100}{\percent}$ in steps of $\SI{0.1}{\percent}$, the normalized root mean square error between Equation~\ref{eq:boundSymUncorrelated} and the equations presented in \cite{kindt:18} for $PI-0M$ is $\SI{1.24}{\percent}$.

\vspace*{2em}
\section{Conclusion}
In this section, we first describe open problems left for future research and then summarize the main results of this paper.
\subsection{Open Problems}
\subsubsection{Problems On Fundamental Limits}
\label{sec:multiDevDisc}
Regarding the future work on fundamental limits, there are two important problems left open. First, what is the lowest latency an asymmetric protocol can guarantee, if the duty-cycles of all devices are unknown? An what is the bound for asymmetric \ac{ND} for duty-cycles for which $\nicefrac{2}{\eta}$ is not an integer?

Second, the bounds derived so far are valid for a pair of devices discovering each other. For unidirectional beaconing, protocols in which $\SI{100}{\percent}$ of all discovery attempts are successful within $L$ time-units can be realized in practice. For increasing numbers of devices discovering each other simultaneously, it is inevitable that their beacons will collide and hence, an increasing number of discovery attempts will fail. Therefore, generalized performance bounds for multi-device scenarios need to be derived. Such bounds are in the form of a function $L(\beta, \gamma, S, P_f)$, which needs to be interpreted as follows. For a given number of devices $S$ with duty-cycles $\beta$ and $\gamma$, in no \ac{ND} protocol, a fraction of at least $1 - P_f$ of all discovery attempts will terminate successfully within less than $L$ time-units. Clearly, for $S \rightarrow 1$ and $P_f \rightarrow 0$, this bound converges to $L$ from Equation~\ref{eq:boundUnidir}. 
The following two mechanisms determine the performance in multi-device scenarios.

\noindent\textbf{1) Lowering the Channel Utilization:} The rate of collisions directly correlates to the channel utilization $\beta$, as described by Equation~\ref{eq:abramsonCollisions}. Hence, devices can reduce the failure probability $P_f$ by reducing $\beta$, which will, however, negatively affect the discovery latencies achieved in the two-device case (cf. Equation~\ref{eq:boundUnidir}).

\noindent\textbf{2) Redundant Coverage:} Optimality in the $L(\beta, \gamma)$ - metric for two devices implies that every initial offset is covered exactly once (cf. Theorems~\ref{thm:1stBeaconingTheorem} and \ref{thm:TcOpt}) and hence, every collision leads to a failed discovery. However, an \ac{ND} protocol might cover multiple or all initial offsets more than once. Hence, for such offsets, more than one beacon would overlap with a reception window, and as long as one of them is not subjected to collisions, the discovery procedure will succeed. Moreover, it seems feasible to construct protocols that first cover every offset exactly once by a beacon sequence $B^\prime$ of length $M$.
In addition, the same offsets are then covered again by concatenations of multiple instances of $B^\prime$. In other words, such protocols would guarantee short latencies in the two-device case while performing potentially optimally also in multi-device scenarios.

The collision of a pair of beacons from two devices often induces an increased collision probability of subsequent pairs of beacons. For example, consider protocols in which beacons are sent with periodic intervals. Since all devices transmit with the same interval, a collision implies that all later beacons will also collide. 
To make protocols robust against failures due to collisions, a beacon schedule needs to fulfill the following property. Given any two beacons that both overlap with a reception window for the same offset $\Phi_1$, their individual collision probabilities should exhibit the lowest possible correlation.
It is currently not clear which degree of such a decorrelation can be actually achieved. Further, measures for decorrelating collision probabilities might reduce the latency performance, because they could prevent beacons from being sent at their optimal points in time. Hence, not all initial offsets can be covered with the fewest possible number of beacons, making additional beacon transmissions necessary.
Besides open questions on decorrelating collisions, for protocols being optimal in the multiple-device case, how many times should every initial offset be covered? These questions need to be studied further in order to derive agnostic bounds in the form of $L(\beta, \gamma, S, P_f)$.
\nomenclature{$P_f$}{Probability of a failed discovery} 

\subsubsection{Problems in Protocol Design}
Our results also outline two important directions for the development of future \ac{ND} protocols. First, there is no existing protocol which, for \textit{every} duty-cycle and every required collision rate, could realize the optimal performance predicted by Theorem~\ref{thm:boundSymChannelConstraint}. 
Second, protocols that contain decorrelation mechanisms to make the collision of each beacon independent from the occurrence of previous collisions have not been studied thoroughly. Though \ac{BLE} applies some random delay for scheduling its beacons \cite{bleSpec50}, the optimal randomization technique to obtain the best trade-off between robustness and worst-case latency remains an open question.

\subsection{Concluding Remarks}
\label{sec:conclusion}
We have presented and proven the correctness of multiple fundamental bounds on the performance of deterministic \ac{ND} protocols. In particular, we have presented bounds for unidirectional beaconing, for symmetric and for asymmetric bi-directional \ac{ND}. Further, we have shown that in the latency/duty-cycle metric, only slotless protocols can reach optimal performance. However, if the channel utilization is constrained, both slotted and slotless protocols can perform optimally.
We have also revealed new important open problems to be addressed by future research.

\vspace*{2em}
\section*{Acknowledgements}
This work was partially supported by the German Research Foundation (DFG) under grant number CH918/5-1 - ``Slotless Neighbor Discovery''.
We gratefully acknowledge Prof. Polly Huang for shepherding our paper at SIGCOMM.

	
	\interlinepenalty=10000
	\bibliographystyle{IEEEtran}
	\bibliography{literature}


\begin{thebibliography}{23}


\ifx \showCODEN    \undefined \def \showCODEN     #1{\unskip}     \fi
\ifx \showDOI      \undefined \def \showDOI       #1{#1}\fi
\ifx \showISBNx    \undefined \def \showISBNx     #1{\unskip}     \fi
\ifx \showISBNxiii \undefined \def \showISBNxiii  #1{\unskip}     \fi
\ifx \showISSN     \undefined \def \showISSN      #1{\unskip}     \fi
\ifx \showLCCN     \undefined \def \showLCCN      #1{\unskip}     \fi
\ifx \shownote     \undefined \def \shownote      #1{#1}          \fi
\ifx \showarticletitle \undefined \def \showarticletitle #1{#1}   \fi
\ifx \showURL      \undefined \def \showURL       {\relax}        \fi
\providecommand\bibfield[2]{#2}
\providecommand\bibinfo[2]{#2}
\providecommand\natexlab[1]{#1}
\providecommand\showeprint[2][]{arXiv:#2}

\bibitem[\protect\citeauthoryear{Abramson}{Abramson}{1970}]%
        {abramson:70}
\bibfield{author}{\bibinfo{person}{N. Abramson}.}
  \bibinfo{year}{1970}\natexlab{}.
\newblock \showarticletitle{THE ALOHA SYSTEM: Another Alternative for Computer
  Communications}. In \bibinfo{booktitle}{{\em ACM fall joint computer
  conference}}.
\newblock


\bibitem[\protect\citeauthoryear{Bakht, Trower, and Kravets}{Bakht
  et~al\mbox{.}}{2012}]%
        {bakht:12}
\bibfield{author}{\bibinfo{person}{M. Bakht}, \bibinfo{person}{M. Trower},
  {and} \bibinfo{person}{R.H. Kravets}.} \bibinfo{year}{2012}\natexlab{}.
\newblock \showarticletitle{Searchlight: Won't You Be My Neighbor?}. In
  \bibinfo{booktitle}{{\em Annual International Conference on Mobile Computing
  and Networking ({MOBICOM})}}. \bibinfo{pages}{185--196}.
\newblock


\bibitem[\protect\citeauthoryear{Barenboim, Dolev, and Ostrovsky}{Barenboim
  et~al\mbox{.}}{2014}]%
        {barenboim:14}
\bibfield{author}{\bibinfo{person}{L. Barenboim}, \bibinfo{person}{S. Dolev},
  {and} \bibinfo{person}{R. Ostrovsky}.} \bibinfo{year}{2014}\natexlab{}.
\newblock \showarticletitle{Deterministic and Energy-Optimal Wireless
  Synchronization}.
\newblock \bibinfo{journal}{{\em ACM Transactions on Sensor Networks (TOSN)\/}}
  \bibinfo{volume}{11}, \bibinfo{number}{1} (\bibinfo{year}{2014}),
  \bibinfo{pages}{13:1--13:25}.
\newblock


\bibitem[\protect\citeauthoryear{Bradonjic, Kohler, and Ostrovsky}{Bradonjic
  et~al\mbox{.}}{2008}]%
        {bradonjic:08}
\bibfield{author}{\bibinfo{person}{M. Bradonjic}, \bibinfo{person}{E. Kohler},
  {and} \bibinfo{person}{R. Ostrovsky}.} \bibinfo{year}{2008}\natexlab{}.
\newblock \showarticletitle{Near-Optimal Radio Use For Wireless Network
  Synchronization}.
\newblock \bibinfo{journal}{{\em CoRR\/}}  \bibinfo{volume}{abs/0810.1756}
  (\bibinfo{year}{2008}).
\newblock


\bibitem[\protect\citeauthoryear{Bradonjic, Kohler, and Ostrovsky}{Bradonjic
  et~al\mbox{.}}{2012}]%
        {bradonjic:12}
\bibfield{author}{\bibinfo{person}{M. Bradonjic}, \bibinfo{person}{E. Kohler},
  {and} \bibinfo{person}{R. Ostrovsky}.} \bibinfo{year}{2012}\natexlab{}.
\newblock \showarticletitle{Near-Optimal Radio Use for Wireless Network
  Synchronization}.
\newblock \bibinfo{journal}{{\em Theoretical Computer Science\/}}
  \bibinfo{volume}{453} (\bibinfo{year}{2012}), \bibinfo{pages}{14--28}.
\newblock


\bibitem[\protect\citeauthoryear{Chen, Fan, Chen, Gerla, Wang, and Li}{Chen
  et~al\mbox{.}}{2015}]%
        {chen:15}
\bibfield{author}{\bibinfo{person}{L. Chen}, \bibinfo{person}{R. Fan},
  \bibinfo{person}{L. Chen}, \bibinfo{person}{M. Gerla}, \bibinfo{person}{T.
  Wang}, {and} \bibinfo{person}{X. Li}.} \bibinfo{year}{2015}\natexlab{}.
\newblock \showarticletitle{On Heterogeneous Neighbor Discovery in Wireless
  Sensor Networks}. In \bibinfo{booktitle}{{\em IEEE Conference on Computer
  Communications (INFOCOM)}}. \bibinfo{pages}{693--701}.
\newblock


\bibitem[\protect\citeauthoryear{Dutta and Culler}{Dutta and Culler}{2008}]%
        {dutta:08}
\bibfield{author}{\bibinfo{person}{P. Dutta} {and} \bibinfo{person}{D.
  Culler}.} \bibinfo{year}{2008}\natexlab{}.
\newblock \showarticletitle{Practical Asynchronous Neighbor Discovery and
  Rendezvous for Mobile Sensing Applications}. In \bibinfo{booktitle}{{\em ACM
  Conference on Embedded Network Sensor Systems (SenSys)}}.
  \bibinfo{pages}{71--84}.
\newblock


\bibitem[\protect\citeauthoryear{Julien, Liu, Murphy, and Picco}{Julien
  et~al\mbox{.}}{2017}]%
        {julien:17}
\bibfield{author}{\bibinfo{person}{C. Julien}, \bibinfo{person}{C. Liu},
  \bibinfo{person}{A.~L. Murphy}, {and} \bibinfo{person}{G.~P. Picco}.}
  \bibinfo{year}{2017}\natexlab{}.
\newblock \showarticletitle{BLEnd: Practical Continuous Neighbor Discovery for
  Bluetooth Low Energy}. In \bibinfo{booktitle}{{\em ACM/IEEE International
  Conference on Information Processing in Sensor Networks (IPSN)}}.
  \bibinfo{pages}{105--116}.
\newblock


\bibitem[\protect\citeauthoryear{Kandhalu, Lakshmanan, and Rajkumar}{Kandhalu
  et~al\mbox{.}}{2010}]%
        {Kandhalu:10}
\bibfield{author}{\bibinfo{person}{A. Kandhalu}, \bibinfo{person}{K.
  Lakshmanan}, {and} \bibinfo{person}{R. Rajkumar}.}
  \bibinfo{year}{2010}\natexlab{}.
\newblock \showarticletitle{{U}-Connect: A Low-Latency Energy-Efficient
  Asynchronous Neighbor Discovery Protocol}. In \bibinfo{booktitle}{{\em
  International Conference on Information Processing in Sensor Networks
  ({IPSN})}}. \bibinfo{pages}{350--361}.
\newblock


\bibitem[\protect\citeauthoryear{Kindt, Saur, and Chakraborty}{Kindt
  et~al\mbox{.}}{2016}]%
        {kindt:18}
\bibfield{author}{\bibinfo{person}{P.H. Kindt}, \bibinfo{person}{M. Saur},
  {and} \bibinfo{person}{S. Chakraborty}.} \bibinfo{year}{2016}\natexlab{}.
\newblock \showarticletitle{Slotless Protocols for Fast and Energy-Efficient
  Neighbor Discovery}.
\newblock \bibinfo{journal}{{\em CoRR\/}}  \bibinfo{volume}{abs/1605.05614}
  (\bibinfo{year}{2016}).
\newblock


\bibitem[\protect\citeauthoryear{Kindt, Saur, and Chakraborty}{Kindt
  et~al\mbox{.}}{2018}]%
        {kindt:15}
\bibfield{author}{\bibinfo{person}{P.H. Kindt}, \bibinfo{person}{M. Saur},
  {and} \bibinfo{person}{S. Chakraborty}.} \bibinfo{year}{2018}\natexlab{}.
\newblock \showarticletitle{Neighbor Discovery Latency in {BLE}-Like
  Protocols}.
\newblock \bibinfo{journal}{{\em {IEEE} Transactions on Mobile Computing
  ({TMC})\/}} \bibinfo{volume}{17}, \bibinfo{number}{3} (\bibinfo{year}{2018}),
  \bibinfo{pages}{617--631}.
\newblock


\bibitem[\protect\citeauthoryear{Kindt, Yunge, Reinerth, and Chakraborty}{Kindt
  et~al\mbox{.}}{2017}]%
        {kindt:17a}
\bibfield{author}{\bibinfo{person}{P.H. Kindt}, \bibinfo{person}{D. Yunge},
  \bibinfo{person}{G. Reinerth}, {and} \bibinfo{person}{S. Chakraborty}.}
  \bibinfo{year}{2017}\natexlab{}.
\newblock \showarticletitle{Griassdi: Mutually Assisted Slotless Neighbor
  Discovery}. In \bibinfo{booktitle}{{\em {ACM/IEEE} International Conference
  on Information Processing in Sensor Networks ({IPSN})}}.
  \bibinfo{pages}{93--104}.
\newblock


\bibitem[\protect\citeauthoryear{\mbox{Bluetooth SIG}}{\mbox{Bluetooth
  SIG}}{2016}]%
        {bleSpec50}
\bibfield{author}{\bibinfo{person}{\mbox{Bluetooth SIG}}.}
  \bibinfo{year}{2016}\natexlab{}.
\newblock \bibinfo{title}{Specification of the {Bluetooth} System 5.0}.
\newblock   (\bibinfo{date}{December} \bibinfo{year}{2016}).
\newblock
\newblock
\shownote{Volume 0, available via \mbox{\url{bluetooth.org}}.}


\bibitem[\protect\citeauthoryear{Meng, Wu, and Chen}{Meng
  et~al\mbox{.}}{2014}]%
        {meng:14}
\bibfield{author}{\bibinfo{person}{T. Meng}, \bibinfo{person}{F. Wu}, {and}
  \bibinfo{person}{G. Chen}.} \bibinfo{year}{2014}\natexlab{}.
\newblock \showarticletitle{On Designing Neighbor Discovery Protocols: A
  Code-Based Approach}. In \bibinfo{booktitle}{{\em IEEE Conference on Computer
  Communications (INFOCOM)}}. \bibinfo{pages}{1689--1697}.
\newblock


\bibitem[\protect\citeauthoryear{Meng, Wu, and Chen}{Meng
  et~al\mbox{.}}{2016}]%
        {meng:16}
\bibfield{author}{\bibinfo{person}{T. Meng}, \bibinfo{person}{F. Wu}, {and}
  \bibinfo{person}{G. Chen}.} \bibinfo{year}{2016}\natexlab{}.
\newblock \showarticletitle{Code-Based Neighbor Discovery Protocols in Mobile
  Wireless Networks}.
\newblock \bibinfo{journal}{{\em {IEEE/ACM} Transactions on Networking
  ({TON})\/}} \bibinfo{volume}{24}, \bibinfo{number}{2} (\bibinfo{year}{2016}),
  \bibinfo{pages}{806--819}.
\newblock


\bibitem[\protect\citeauthoryear{Portal}{Portal}{2018}]%
        {statista_BLE:2018}
\bibfield{author}{\bibinfo{person}{Statista The~Statistics Portal}.}
  \bibinfo{year}{2018}\natexlab{}.
\newblock \bibinfo{title}{{Bluetooth Low Energy} ({BLE}) Enabled Devices Market
  Volume Worldwide, from 2013 to 2020 (in Million Units).}
\newblock   (\bibinfo{year}{2018}).
\newblock
\newblock
\shownote{\url{www.statista.com/statistics/750569/worldwide-bluetooth-low-energy-device-market-volume}.}


\bibitem[\protect\citeauthoryear{Qiu, Li, Xu, and Li}{Qiu
  et~al\mbox{.}}{2016}]%
        {qiu:16}
\bibfield{author}{\bibinfo{person}{Y. Qiu}, \bibinfo{person}{S. Li},
  \bibinfo{person}{X. Xu}, {and} \bibinfo{person}{Z. Li}.}
  \bibinfo{year}{2016}\natexlab{}.
\newblock \showarticletitle{Talk More Listen Less: Energy-Efficient Neighbor
  Discovery in Wireless Sensor Networks}. In \bibinfo{booktitle}{{\em IEEE
  Conference on Computer Communications (INFOCOM)}}. \bibinfo{pages}{1--9}.
\newblock


\bibitem[\protect\citeauthoryear{Schurgers, Tsiatsis, Ganeriwal, and
  Srivastava}{Schurgers et~al\mbox{.}}{2002}]%
        {schurgers:02}
\bibfield{author}{\bibinfo{person}{C. Schurgers}, \bibinfo{person}{V.
  Tsiatsis}, \bibinfo{person}{S. Ganeriwal}, {and} \bibinfo{person}{M.
  Srivastava}.} \bibinfo{year}{2002}\natexlab{}.
\newblock \showarticletitle{Optimizing Sensor Networks in the
  Energy-Latency-Density Design Space}.
\newblock \bibinfo{journal}{{\em IEEE Transactions on Mobile Computing
  (TMC)\/}} \bibinfo{volume}{1}, \bibinfo{number}{1} (\bibinfo{year}{2002}),
  \bibinfo{pages}{70--80}.
\newblock
\showISSN{1536-1233}


\bibitem[\protect\citeauthoryear{Sun, Yang, Keyu, and Yunhao}{Sun
  et~al\mbox{.}}{2014}]%
        {Sun:14}
\bibfield{author}{\bibinfo{person}{W. Sun}, \bibinfo{person}{Z. Yang},
  \bibinfo{person}{W. Keyu}, {and} \bibinfo{person}{L. Yunhao}.}
  \bibinfo{year}{2014}\natexlab{}.
\newblock \showarticletitle{Hello: A Generic Flexible Protocol for Neighbor
  Discovery}. In \bibinfo{booktitle}{{\em IEEE Conference on Computer
  Communications (INFOCOM)}}. \bibinfo{pages}{540--548}.
\newblock


\bibitem[\protect\citeauthoryear{Tseng, Hsu, and Hsieh}{Tseng
  et~al\mbox{.}}{2002}]%
        {Tseng:02}
\bibfield{author}{\bibinfo{person}{Y.C. Tseng}, \bibinfo{person}{C.-S. Hsu},
  {and} \bibinfo{person}{T.-Y. Hsieh}.} \bibinfo{year}{2002}\natexlab{}.
\newblock \showarticletitle{Power-Saving Protocols for {IEEE} 802.11 Based
  Multi-Hop Ad Hoc Networks}. In \bibinfo{booktitle}{{\em IEEE Conference on
  Computer Communications (INFOCOM)}}.
\newblock


\bibitem[\protect\citeauthoryear{Zhang, He, Liu, Gu, Ye, Ganti, and Lei}{Zhang
  et~al\mbox{.}}{2012}]%
        {zhang:12}
\bibfield{author}{\bibinfo{person}{D. Zhang}, \bibinfo{person}{T. He},
  \bibinfo{person}{Y. Liu}, \bibinfo{person}{Y. Gu}, \bibinfo{person}{F. Ye},
  \bibinfo{person}{R.~K. Ganti}, {and} \bibinfo{person}{H. Lei}.}
  \bibinfo{year}{2012}\natexlab{}.
\newblock \showarticletitle{\emph{Acc}: Generic On-Demand Accelerations for
  Neighbor Discovery in Mobile Applications}. In \bibinfo{booktitle}{{\em {ACM}
  Conference on Embedded Network Sensor Systems, (SenSys)}}.
\newblock


\bibitem[\protect\citeauthoryear{Zheng, Hou, and Sha}{Zheng
  et~al\mbox{.}}{2006}]%
        {zheng:06}
\bibfield{author}{\bibinfo{person}{R. Zheng}, \bibinfo{person}{J.C. Hou}, {and}
  \bibinfo{person}{L. Sha}.} \bibinfo{year}{2006}\natexlab{}.
\newblock \showarticletitle{Optimal Block Design for Asynchronous Wake-Up
  Schedules and Its Applications in Multihop Wireless Networks}.
\newblock \bibinfo{journal}{{\em {IEEE} Transactions on Mobile Computing
  (TMC)\/}} \bibinfo{volume}{5}, \bibinfo{number}{9} (\bibinfo{year}{2006}),
  \bibinfo{pages}{1228--1241}.
\newblock


\bibitem[\protect\citeauthoryear{Zheng, Hou, and Sha}{Zheng
  et~al\mbox{.}}{2003}]%
        {zheng:03}
\bibfield{author}{\bibinfo{person}{R. Zheng}, \bibinfo{person}{J.~C. Hou},
  {and} \bibinfo{person}{L. Sha}.} \bibinfo{year}{2003}\natexlab{}.
\newblock \showarticletitle{Asynchronous Wakeup for Ad Hoc Networks}. In
  \bibinfo{booktitle}{{\em ACM International Symposium on Mobile Ad Hoc
  Networking \& Computing (MobiHoc)}}.
\newblock


\end{thebibliography}


\begin{thebibliography}{10}
\providecommand{\url}[1]{#1}
\csname url@samestyle\endcsname
\providecommand{\newblock}{\relax}
\providecommand{\bibinfo}[2]{#2}
\providecommand{\BIBentrySTDinterwordspacing}{\spaceskip=0pt\relax}
\providecommand{\BIBentryALTinterwordstretchfactor}{4}
\providecommand{\BIBentryALTinterwordspacing}{\spaceskip=\fontdimen2\font plus
\BIBentryALTinterwordstretchfactor\fontdimen3\font minus
  \fontdimen4\font\relax}
\providecommand{\BIBforeignlanguage}[2]{{%
\expandafter\ifx\csname l@#1\endcsname\relax
\typeout{** WARNING: IEEEtran.bst: No hyphenation pattern has been}%
\typeout{** loaded for the language `#1'. Using the pattern for}%
\typeout{** the default language instead.}%
\else
\language=\csname l@#1\endcsname
\fi
#2}}
\providecommand{\BIBdecl}{\relax}
\BIBdecl

\bibitem{mcglynn:02}
M.~J. McGlynn and S.~A. Borbash, ``Birthday protocols for low energy deployment
  and flexible neighbor discovery in ad hoc wireless networks,'' in \emph{ACM
  International Symposium on Mobile Ad Hoc Networking \& Computing (MobiHoc)},
  2001.

\bibitem{margolies:16}
R.~Margolies, G.~Grebla, T.~Chen, D.~Rubenstein, and G.~Zussman, ``Panda:
  Neighbor discovery on a power harvesting budget,'' \emph{{IEEE} Journal on
  Selected Areas in Communications}, vol.~34, no.~12, pp. 3606--3619, 2016.

\bibitem{you:11}
L.~You, Z.~Yuan, P.~Yang, and G.~Chen, ``Aloha-like neighbor discovery in
  low-duty-cycle wireless sensor networks,'' in \emph{IEEE Wireless
  Communications and Networking Conference ({WCNC})}, 2011.

\bibitem{vasudevan:2009}
S.~Vasudevan, D.~Towsley, D.~Goeckel, and R.~Khalili, ``Neighbor discovery in
  wireless networks and the coupon collector's problem,'' in \emph{Annual
  International Conference on Mobile Computing and Networking ({MobiCom})},
  2009.

\bibitem{schurgers:02}
C.~Schurgers, V.~Tsiatsis, S.~Ganeriwal, and M.~Srivastava, ``Optimizing sensor
  networks in the energy-latency-density design space,'' \emph{IEEE
  Transactions on Mobile Computing (TMC)}, vol.~1, no.~1, pp. 70--80, 2002.

\bibitem{Tseng:02}
Y.~Tseng, C.-S. Hsu, and T.-Y. Hsieh, ``Power-saving protocols for {IEEE}
  802.11 based multi-hop ad hoc networks,'' in \emph{IEEE Conference on
  Computer Communications (INFOCOM)}, 2002.

\bibitem{dutta:08}
P.~Dutta and D.~Culler, ``Practical asynchronous neighbor discovery and
  rendezvous for mobile sensing applications,'' in \emph{ACM Conference on
  Embedded Network Sensor Systems (SenSys)}, 2008, pp. 71--84.

\bibitem{Kandhalu:10}
A.~Kandhalu, K.~Lakshmanan, and R.~Rajkumar, ``{U}-connect: A low-latency
  energy-efficient asynchronous neighbor discovery protocol,'' in
  \emph{International Conference on Information Processing in Sensor Networks
  ({IPSN})}, 2010, pp. 350--361.

\bibitem{bakht:12}
M.~Bakht, M.~Trower, and R.~Kravets, ``Searchlight: Won't you be my neighbor?''
  in \emph{Annual International Conference on Mobile Computing and Networking
  ({MOBICOM})}, 2012, pp. 185--196.

\bibitem{meng:14}
T.~Meng, F.~Wu, and G.~Chen, ``On designing neighbor discovery protocols: A
  code-based approach,'' in \emph{IEEE Conference on Computer Communications
  (INFOCOM)}, 2014, pp. 1689--1697.

\bibitem{meng:16}
------, ``Code-based neighbor discovery protocols in mobile wireless
  networks,'' \emph{{IEEE/ACM} Transactions on Networking ({TON})}, vol.~24,
  no.~2, pp. 806--819, 2016.

\bibitem{chen:15}
L.~Chen, R.~Fan, L.~Chen, M.~Gerla, T.~Wang, and X.~Li, ``On heterogeneous
  neighbor discovery in wireless sensor networks,'' in \emph{IEEE Conference on
  Computer Communications (INFOCOM)}, 2015, pp. 693--701.

\bibitem{zhang:12}
D.~Zhang, T.~He, Y.~Liu, Y.~Gu, F.~Ye, R.~K. Ganti, and H.~Lei, ``\emph{Acc}:
  Generic on-demand accelerations for neighbor discovery in mobile
  applications,'' in \emph{{ACM} Conference on Embedded Network Sensor Systems,
  (SenSys)}, 2012.

\bibitem{Sun:14}
W.~Sun, Z.~Yang, W.~Keyu, and L.~Yunhao, ``Hello: A generic flexible protocol
  for neighbor discovery,'' in \emph{IEEE Conference on Computer Communications
  (INFOCOM)}, 2014, pp. 540--548.

\bibitem{qiu:16}
Y.~Qiu, S.~Li, X.~Xu, and Z.~Li, ``Talk more listen less: Energy-efficient
  neighbor discovery in wireless sensor networks,'' in \emph{IEEE Conference on
  Computer Communications (INFOCOM)}, 2016, pp. 1--9.

\bibitem{julien:17}
C.~Julien, C.~Liu, A.~L. Murphy, and G.~P. Picco, ``Blend: Practical continuous
  neighbor discovery for bluetooth low energy,'' in \emph{ACM/IEEE
  International Conference on Information Processing in Sensor Networks
  (IPSN)}, 2017, pp. 105--116.

\bibitem{kindt:17a}
P.~Kindt, D.~Yunge, G.~Reinerth, and S.~Chakraborty, ``Griassdi: Mutually
  assisted slotless neighbor discovery,'' in \emph{{ACM/IEEE} International
  Conference on Information Processing in Sensor Networks ({IPSN})}, 2017, pp.
  93--104.

\bibitem{kindt:18}
P.~Kindt, M.~Saur, and S.~Chakraborty, ``Slotless protocols for fast and
  energy-efficient neighbor discovery,'' \emph{CoRR}, vol. abs/1605.05614,
  2016.

\bibitem{wei:16}
L.~Wei, B.~Zhou, X.~Ma, D.~Chen, J.~Zhang, J.~Peng, Q.~Luo, L.~Sun, D.~Li, and
  L.~Chen, ``Lightning: A high-efficient neighbor discovery protocol for low
  duty cycle wsns,'' \emph{IEEE Communications Letters}, vol.~20, no.~5, pp.
  966--969, 2016.

\bibitem{zheng:03}
R.~Zheng, J.~C. Hou, and L.~Sha, ``Asynchronous wakeup for ad hoc networks,''
  in \emph{ACM International Symposium on Mobile Ad Hoc Networking \& Computing
  (MobiHoc)}, 2003.

\bibitem{zheng:06}
R.~Zheng, J.~Hou, and L.~Sha, ``Optimal block design for asynchronous wake-up
  schedules and its applications in multihop wireless networks,'' \emph{{IEEE}
  Transactions on Mobile Computing (TMC)}, vol.~5, no.~9, pp. 1228--1241, 2006.

\bibitem{chen:18}
T.~Chen, J.~Ghaderi, D.~Rubenstein, and G.~Zussman, ``Maximizing broadcast
  throughput under ultra-low-power constraints,'' \emph{{IEEE/ACM} Transactions
  on Networking (TON)}, vol.~26, no.~2, pp. 779--792, 2018.

\bibitem{kandhalu:13}
A.~Kandhalu, A.~Xhafa, and S.~Hosur, ``Towards bounded-latency bluetooth low
  energy for in-vehicle network cable replacement,'' in \emph{International
  Conference on Connected Vehicles and Expo (ICCVE)}, 2013, pp. 635--640.

\bibitem{lee:19}
W.~Lee, J.~Youn, and T.~Song, ``Prime-number-assisted block-based neighbor
  discovery protocol in wireless sensor networks,'' \emph{International Journal
  of Distributed Sensor Networks}, vol.~15, no.~1, p. 1550147719826240, 2019.

\bibitem{kim:13}
K.~Kim, H.~Roh, W.~Lee, S.~Lee, and D.-Z. Du, ``{PND}: a p-persistent neighbor
  discovery protocol in wireless networks,'' \emph{Wireless Communications and
  Mobile Computing}, vol.~13, no.~7, pp. 650--662, 2013.

\bibitem{vasudevan:13}
S.~Vasudevan, M.~Adler, D.~Goeckel, and D.~Towsley, ``Efficient algorithms for
  neighbor discovery in wireless networks,'' \emph{IEEE/ACM Transactions on
  Networking ({TON})}, vol.~21, no.~1, pp. 69--83, 2013.

\bibitem{liu:11}
A.~Purohit, B.~Priyantha, and J.~Liu, ``Wiflock: Collaborative group discovery
  and maintenance in mobile sensor networks,'' in \emph{{ACM/IEEE}
  International Conference on Information Processing in Sensor Networks
  ({IPSN})}, 2011.

\bibitem{vasudevan:05}
S.~Vasudevan and D.~Kurose, J.and~Towsley, ``On neighbor discovery in wireless
  networks with directional antennas,'' in \emph{{IEEE} Conference on Computer
  Communications ({INFOCOM})}, vol.~4, 2005.

\bibitem{jakllari:07}
G.~Jakllari, W.~Luo, and S.~V. Krishnamurthy, ``An integrated neighbor
  discovery and {MAC} protocol for ad hoc networks using directional
  antennas,'' \emph{IEEE Transactions on Wireless Communications ({TWC})},
  vol.~6, no.~3, pp. 1114--1024, 2007.

\bibitem{zhang:08}
Z.~Zhang and B.~Li, ``Neighbor discovery in mobile ad hoc self-configuring
  networks with directional antennas: algorithms and comparisons,'' \emph{IEEE
  Transactions on Wireless Communications ({TWC})}, vol.~7, no.~5, pp.
  1540--1549, 2008.

\bibitem{karowski:11}
N.~Karowski, A.~C. Viana, and A.~Wolisz, ``Optimized asynchronous multi-channel
  neighbor discovery,'' in \emph{IEEE Conference on Computer Communications
  ({INFOCOM})}, 2011.

\bibitem{zeng:11}
W.~Zeng, S.~Vasudevan, X.~Chen, B.~Wang, A.~Russell, and W.~Wei, ``Neighbor
  discovery in wireless networks with multipacket reception,'' in
  \emph{Symposium on Mobile Ad Hoc Networking and Computing ({MobiHoc})}, 2011.

\bibitem{borbash:16}
S.~Borbash, A.~Ephremides, and M.~J. McGlynn, ``An asynchronous neighbor
  discovery algorithm for wireless sensor networks,'' \emph{Ad Hoc Networks},
  vol.~5, no.~7, pp. 998--1016, 2007.

\bibitem{wang_2:13}
K.~Wang, X.~Mao, and L.~Y., ``{BlindDate}: A neighbor discovery protocol,'' in
  \emph{International Conference on Parallel Processing ({ICCP})}, 2013, pp.
  120--129.

\bibitem{zhang:17}
Y.~Zhang, K.~Bian, L.~Chen, P.~Zhou, and X.~Li, ``Dynamic slot-length control
  for reducing neighbor discovery latency in wireless sensor networks,'' in
  \emph{IEEE Global Communications Conference (GLOBECOM)}, 2017.

\bibitem{guo:17}
X.~Guo, B.~Chen, and M.~Chan, ``Analysis and design of low-duty protocol for
  smartphone neighbor discovery,'' \emph{IEEE Transactions on Mobile Computing
  (TMC)}, vol.~16, no.~12, pp. 3294--3307, 2017.

\bibitem{zhang:17_b}
D.~Zhang, T.~He, F.~Ye, R.~K. Ganti, and H.~Lei, ``Neighbor discovery and
  rendezvous maintenance with extended quorum systems for mobile
  applications,'' \emph{IEEE Transactions on Mobile Computing (TMC)}, vol.~16,
  no.~7, pp. 1967--1980, 2017.

\bibitem{cao:18}
Z.~Cao, Z.~Gu, Y.~Wang, and H.~Cui, ``Panacea: A low-latency, energy-efficient
  neighbor discovery protocol for wireless sensor networks,'' in \emph{IEEE
  Wireless Communications and Networking Conference (WCNC)}, 2018.

\bibitem{chen:17}
H.~Chen, W.~Lou, Z.~Wang, and F.~Xia, ``On achieving asynchronous
  energy-efficient neighbor discovery for mobile sensor networks,'' \emph{IEEE
  Transactions on Emerging Topics in Computing (TETC)}, 2017.

\bibitem{chen:16_b}
L.~Chen, K.~Bian, and M.~Zheng, ``Never live without neighbors: From single- to
  multi-channel neighbor discovery for mobile sensing applications,''
  \emph{IEEE/ACM Transactions on Networking (TON)}, vol.~24, no.~5, pp.
  3148--3161, 2016.

\bibitem{lee:15}
W.~Lee, S.~Choi, N.~Kim, J.~H. Youn, and D.~Moore, ``Block combination
  selection scheme for neighbor discovery protocol,'' in \emph{International
  Conference on Communication Systems and Network Technologies (CSNT)}, 2015.

\bibitem{yang:15}
D.~Yang, J.~Shin, J.~Kim, and G.~H. Kim, ``Opeed: Optimal energy-efficient
  neighbor discovery scheme in opportunistic networks,'' \emph{Journal of
  Communications and Networks}, vol.~17, no.~1, pp. 34--39, 2015.

\bibitem{wang:14}
H.~Wang, J.~Ma, Y.~Liu, W.~Liu, and L.~Wang, ``Bi-directional probing for
  neighbor discovery,'' in \emph{IEEE International Conference on Computational
  Science and Engineering (CSE)}, 2014.

\bibitem{hess:14}
A.~Hess, E.~Hyytiä, and J.~Ott, ``Efficient neighbor discovery in mobile
  opportunistic networking using mobility awareness,'' in \emph{International
  Conference on Communication Systems and Networks (COMSNETS)}, 2014.

\bibitem{zhang:13}
M.~Zhang, L.~Zhang, P.~Yang, and Y.~Yan, ``{McDisc}: A reliable neighbor
  discovery protocol in low duty cycle and multi-channel wireless networks,''
  in \emph{IEEE International Conference on Networking, Architecture and
  Storage (NAS)}, 2013.

\bibitem{li:13}
B.~Li, W.~Feng, L.~Zhang, and C.~Spanos, ``{DEPEND}: Density adaptive power
  efficient neighbor discovery for wearable body sensors,'' in \emph{IEEE
  International Conference on Automation Science and Engineering (CASE)}, 2013.

\bibitem{cohen:11}
R.~Cohen and B.~Kapchits, ``Continuous neighbor discovery in asynchronous
  sensor networks,'' \emph{IEEE/ACM Transactions on Networking (TON)}, vol.~19,
  no.~1, pp. 69--79, 2011.

\bibitem{yang:09}
D.~Yang, J.~Shin, J.~Kim, and C.~Kim, ``An energy-optimal scheme for neighbor
  discovery in opportunistic networking,'' in \emph{IEEE Consumer
  Communications and Networking Conference (CCNC)}, 2009.

\bibitem{statista_BLE:2018}
S.~T.~S. Portal, ``{Bluetooth Low Energy} ({BLE}) enabled devices market volume
  worldwide, from 2013 to 2020 (in million units).'' 2018,
  \url{www.statista.com/statistics/750569/worldwide-bluetooth-low-energy-device-market-volume}.

\bibitem{BleFindMeProf}
\mbox{Bluetooth SIG}, ``Find me profile specificiation,'' June 2011, revision
  V10r00, available via \mbox{\url{bluetooth.org}}.

\bibitem{kindt:15}
P.~Kindt, M.~Saur, and S.~Chakraborty, ``Neighbor discovery latency in
  {BLE}-like protocols,'' \emph{{IEEE} Transactions on Mobile Computing
  ({TMC})}, vol.~17, no.~3, pp. 617--631, 2018.

\bibitem{bradonjic:12}
M.~Bradonjic, E.~Kohler, and R.~Ostrovsky, ``Near-optimal radio use for
  wireless network synchronization,'' \emph{Theoretical Computer Science}, vol.
  453, pp. 14--28, 2012.

\bibitem{barenboim:14}
L.~Barenboim, S.~Dolev, and R.~Ostrovsky, ``Deterministic and energy-optimal
  wireless synchronization,'' \emph{ACM Transactions on Sensor Networks
  (TOSN)}, vol.~11, no.~1, pp. 13:1--13:25, 2014.

\bibitem{abramson:70}
N.~Abramson, ``The aloha system: Another alternative for computer
  communications,'' in \emph{ACM fall joint computer conference}, 1970.

\bibitem{nrf51822:14}
\mbox{Nordic Semiconductor {ASA}}, ``{nRF51822} product spec. v3.1,'' 2014,
  available via \mbox{\url{nordicsemi.com}}.

\bibitem{chen:16}
L.~Chen and K.~Bian, ``Neighbor discovery in mobile sensing applications,''
  \emph{Ad Hoc Networks}, vol.~48, no.~C, pp. 38--52, 2016.

\bibitem{bleSpec50}
\mbox{Bluetooth SIG}, ``Specification of the {Bluetooth} system 5.0,'' December
  2016, volume 0, available via \mbox{\url{bluetooth.org}}.

\end{thebibliography}
	\begin{appendix}
		\section*{Appendices}
		\section{Non-Repetitive Reception Window Sequences}
\label{app:non_repetitive_beacon_sequences}
Throughout this paper, we have restricted our considerations to infinite length reception window sequences $C_\infty$ that are given by concatenations of some finite sequence $C$. Though all currently known deterministic \ac{ND} protocols are constructed accordingly, reception window sequences that continuously alter over time are also feasible.
In what follows, we study such sequences and establish that all our presented bounds remain valid for them.

Let us consider an arbitrary pattern of reception windows of infinite length $C_\infty$. Such a $C_\infty$ is characterized by its reception duty-cycle $\gamma$. As in Section~\ref{sec:analysis}, we consider a sequence $B^\prime$ that consists of those beacons that are sent after both devices have come into range. Obviously, the first beacon $b_1 \in B^\prime$ is received successfully if it directly overlaps with one of the reception windows. The fraction of time-units at which a transmission of $b_1$ leads to a reception is therefore identical to $\gamma$. Another beacon that is sent by $\lambda_1$ time units later leads to additional points in time at which $b_1$ can be sent, such that one beacon out of $b_1, b_2$ is received successfully. These additional points in time lie $\lambda_1$ time-units earlier. Hence, like in Section~\ref{sec:analysis}, such points in time for later beacons are given by translating those of earlier ones to the left.
If every point in time is covered by exactly one such translation, the tuple ($B^\prime, C_\infty$) is disjoint and deterministic, and hence potentially optimal. This holds also true for cases in which $C_\infty$ is not an infinite concatenation of the same $C$. The number of beacons $M$ that need to be sent for guaranteeing deterministic discovery is therefore identical to the number of translations of the reception pattern $C_\infty$, such that every point in time overlaps with exactly one such translation. It is:

\begin{equation}
\label{eq:BeaconingTheoremNonRep}
M = \left\lceil \frac{1}{\gamma} \right\rceil
\end{equation}
This is identical to Theorem~\ref{thm:1stBeaconingTheorem}, and hence all bounds remain unchanged.
		\section{Implications of Same Sequences on Both Devices}
\label{app:same_sequences}
Throughout the paper, we have assumed that $C_\infty$ does not impose any constraints on scheduling the beacons in $B_\infty$ on the same device. In this section, we study the relaxation of this assumption.

\subsection{Symmetric Sequences}
We first study the case in which both devices \textit{E} and \textit{F} run the same tuple of sequences ($B_\infty$, $C_\infty$). Here, $B_\infty$ is designed such that a beacon overlap with $C_\infty$ is guaranteed for all initial offsets. Hence, not only an overlap of a beacon of $B_F$ with $C_{E,\infty}$ is guaranteed, but also an overlap of a beacon of $B_E$ with $C_{E,\infty}$.
Such an overlap implies that the affected reception window needs to be interrupted for a certain amount of time.

For an ideal radio (i.e., a radio that does not require any time to switch from reception to transmission and vice-versa, see Section~\ref{app:radio_overheads}), this amount of time is identical to one beacon transmission duration $\omega$. A beacon sent by another device within this period of time would collide and therefore would not be received successfully, even if the radio was able to receive and transmit simultaneously. 

However, a real-world radio needs a certain amount of time $d_{oTxRx}$ to switch from transmission to reception and an overhead $d_{oRxTx}$ to switch from reception to transmission, during which no communication can be carried out. We in the following analyze the impact of this.
Towards this, we next compute the time-fraction of all reception windows in $C_\infty$, during which the radio is unable to receive.
\nomenclature{$d_{oRxTx}$}{Effective additional active time for switching from reception to transmission}
\nomenclature{$d_{oTxRx}$}{Effective additional active time for switching from transmission to reception}

Since an optimal tuple of sequences ($C_{\infty}$, $B^\prime)$ is designed such that every initial offset is covered exactly once, exactly one beacon of $B^\prime$ will overlap with a reception window for every possible initial offset. For every such overlap, the radio is unable to receive incoming beacons for $d_{oTxRx} + d_{oRxTx} + \omega$ time-units within the affected reception windows.

In a tuple $B_\infty, C_\infty$, how frequent do such overlaps occur and which fraction of the total reception time is ``blocked'' by them?
In optimal protocols, exactly one beacon overlaps with a reception window per worst-case latency $L$ (cf. Section~\ref{sec:uniDirBeaconing}). From Theorem~\ref{eq:dmOptUnoptimized} follows that for optimal values of $\gamma$, $L = T_C \cdot \nicefrac{1}{(\sum_{i=1}^{n_C} c_i)} \cdot \nicefrac{\omega}{\beta}$, and hence $L$ is always divisible by $T_C$. In every instance of $T_C$, there are $\sum_{i=1}^{n_C} d_i$ time-units during which the radio is scanning, and therefore, the radio spends $\nicefrac{L}{T_c} \cdot \sum_{i=1}^{n_C} d_i = \nicefrac{\omega}{\beta}$ time-units per worst-case latency $L$ for scanning.
The probability of failed discoveries is identical to the fraction of ``blocked'' time per $L$, which leads to the following equation.
\begin{equation}
P_{fail} = \frac{\beta}{\omega} \cdot (d_{oTxRx} + d_{oRxTx} + \omega)
\end{equation} 
In this equation, we assume that the amount of time during which the radio is ``blocked'' per beacon that overlaps with a reception window of the same device is always identical to $d_{oTxRx} + d_{oRxTx} + \omega$ time-units. We in the following prove this assumption.

\begin{figure}[tb]
	\centering
	\includegraphics[width=\imgWidth]{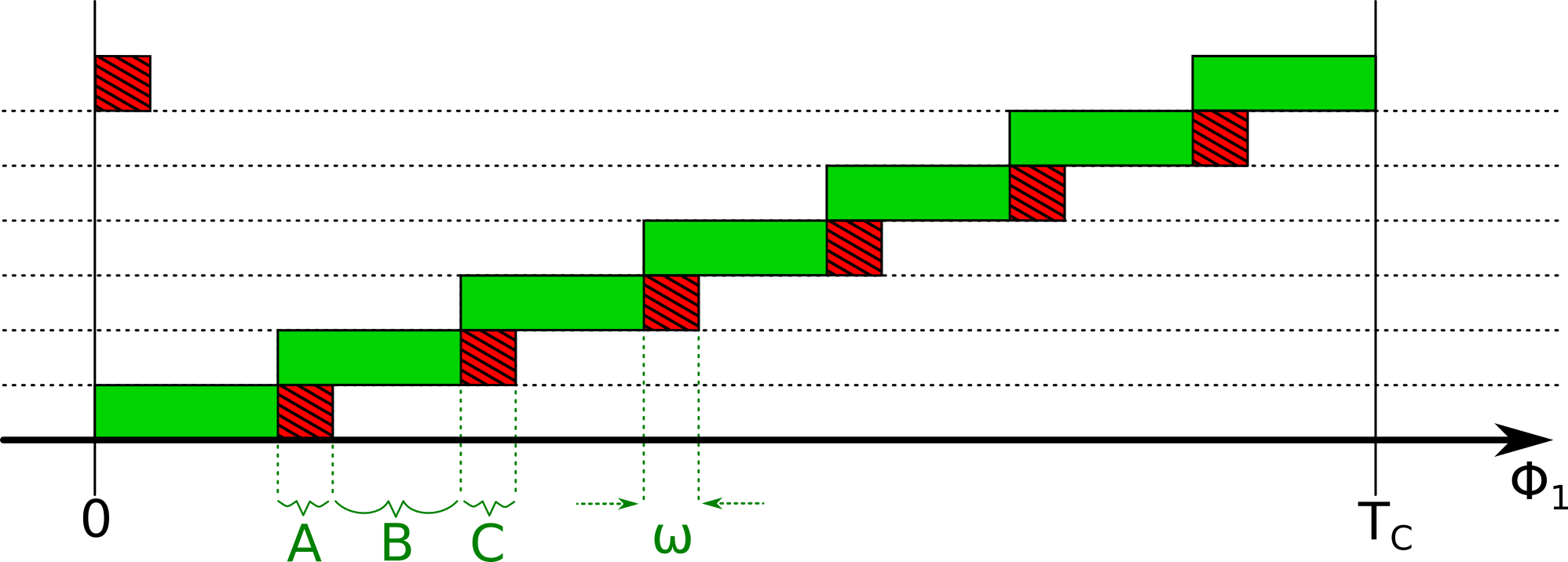}
	\caption{Coverage map of a deterministic beacon sequence $B^\prime_{F}$ in conjunction with a certain $C_{E,\infty}$. The offsets covered by any reception window are composed by a Part $\mathcal{A}$ that overlaps with the last $\omega$ time-units of another reception window, a Part $\mathcal{B}$ that is disjoint and a Part $\mathcal{C}$, during which an incoming beacon is not successfully received.}
	\label{fig:coverageMapOverlapping} 
\end{figure}

Recall from Section~\ref{sec:coverage_determinism} that every beacon of a deterministic Sequence $B^\prime$, in conjunction with a reception window from $C_\infty$ of a remote device, leads to a certain contiguous range of covered offsets, which we in the following call a \textit{coverage image}. If the initial offset $\Phi_1$ lies within one of these coverage images, $B^\prime$ is received successfully.
Figure~\ref{fig:coverageMapOverlapping} exemplifies a coverage map of a non-redundant and deterministic (and hence potentially optimal) \ac{ND} protocol. Here, $C \in C_\infty$ consists of only one reception window and hence, there is one coverage image per beacon. Recall that if a remote device sends a beacon during the last $\omega$ time-units of every scan window, it is not received successfully (cf. Section~\ref{app:non_zero_length_packets}). We can therefore subdivide every coverage image of an optimal protocol into the following three parts $\mathcal{A}$, $\mathcal{B}$ and $\mathcal{C}$ (cf. Figure~\ref{fig:coverageMapOverlapping}).
\begin{itemize}
	\item Part $\mathcal{C}$ has a length of $\omega$ time-units, and a beacon of the remote device that falls into this part will not be received successfully. Therefore, such Parts $\mathcal{C}$ do not contribute to the overall coverage.
	\item To nevertheless ensure discovery if a beacon falls into such a Part $\mathcal{C}$ of a coverage image, each Part $\mathcal{C}$ is also covered by the Part $\mathcal{A}$ of another coverage image, which also has a length of $\omega$ time-units. 
	\item The remaining part \textit{B} is disjoint, i.e., no part of any other coverage image overlaps with it.
\end{itemize}
On a device \textit{E}, we know that exactly one beacon of $B_{E,\infty}$ will overlap with at least one reception window of $C_{E,\infty}$ per $L$, which effectively interrupts or shortens the affected scan window. Such an overlap could happen in one of the following three ways.
\begin{enumerate}
	\item The overlapping beacon falls into Part~$\mathcal{B}$, such that
	a contiguous duration of $d_{oTxRx} + d_{oRxTx} + \omega$ time-units is blocked (e.g., it falls into the center of Part $\mathcal{B}$). 
	\item The overlapping beacon falls into the beginning (e.g., Part~$\mathcal{A}$)
	of the scan window. Therefore, the ``blocked'' amount of time would also overlap with the neighboring Part~$\mathcal{B}$ (cf. Figure~\ref{fig:coverageMapOverlapping}).
	Hence, the amount of occupied scanning time is equal to is $d_{oTxRx} + d_{oRxTx} + \omega$ also for this situation.
	\item The same holds true for a beacon falling into the end of the scan window (e.g., into Part~$\mathcal{C}$), where parts of the ``blocked'' amount of time overlap with a Part~$\mathcal{A}$ and possibly also $\mathcal{B}$ of another scan window.
\end{enumerate}
Hence, in all three cases, the amount of ``blocked'' time is $d_{oTxRx} + d_{oRxTx} + \omega$. 

\subsection{Asymmetric Sequences}
\begin{figure}[tb]
	\centering
	\includegraphics[width=\imgWidth]{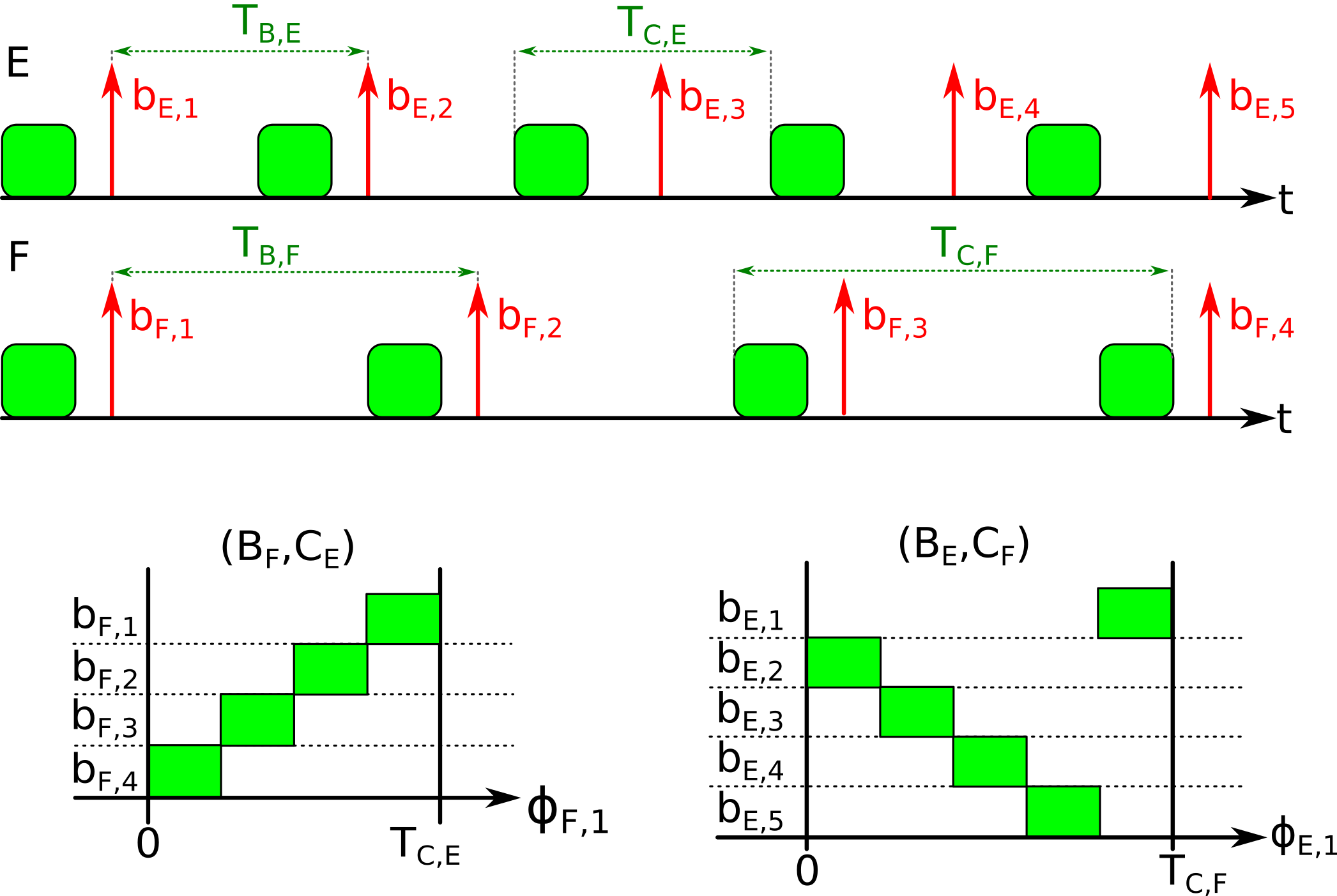}
	\caption{Asymmetric sequences and their coverage maps.}
	\label{fig:nonOverlappingAsymSchedules} 
\end{figure}

For asymmetric discovery (i.e., both devices have different duty-cycles), a quadruple of beacon- and reception window sequences can be designed such that $B_\infty$ and $C_\infty$ on the same device never overlap, while allowing for optimal (i.e., disjoint coverage) and deterministic two-way discovery between the two devices. Figure~\ref{fig:nonOverlappingAsymSchedules} depicts a pair of tuples $(B_{F,\infty}, C_{F,\infty})$ and $(B_{E,\infty}, C_{E,\infty})$ along with the corresponding coverage maps. As can be seen, $(B_{F,\infty}, C_{E,\infty})$ and $(B_{E,\infty}, C_{F,\infty})$ realize disjoint and deterministic discovery, while the sequences on the same device never overlap. 

		\section{Mutual Exclusive Unidirectional Discovery}
\label{sec:correlatedSchedules}
\begin{figure}[bt]
	\centering
	\includegraphics[width=\imgWidth]{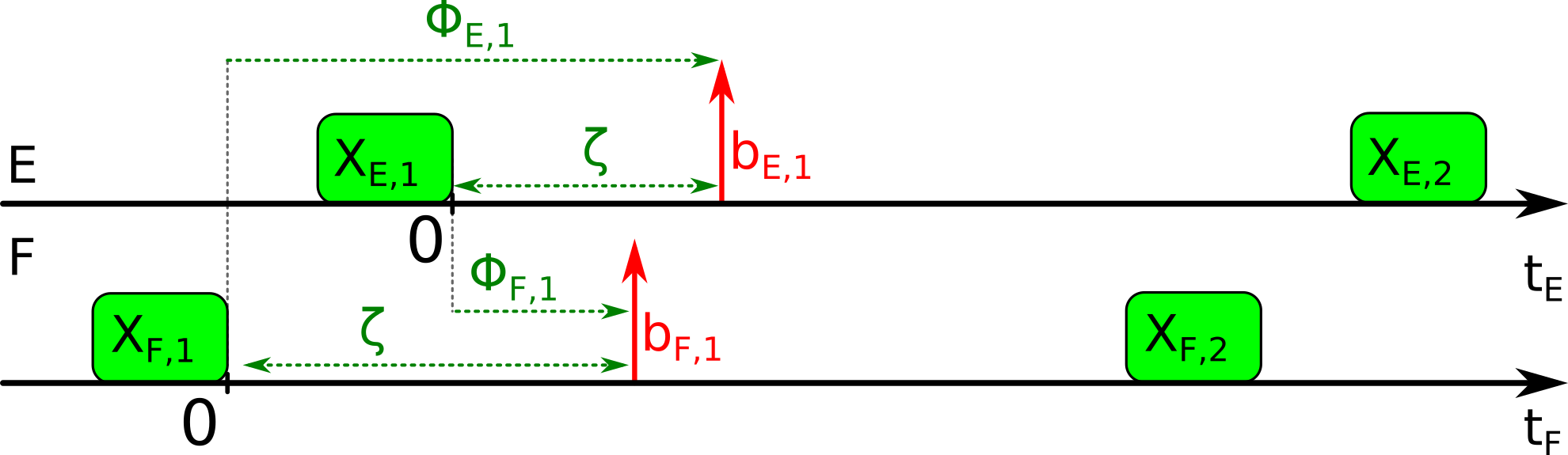}
	\caption{Correlated offsets $\Phi_{F,1}$ and $\Phi_{E,1}$ in the sequences of two devices \textit{E} and \textit{F}.}
	\label{fig:crossCorrelation} 
\end{figure}
In Section~\ref{sec:uniDirBeaconing}, we have studied unidirectional discovery in the sense that one device~\textit{F} could discover \textit{E} without \textit{E} discovering \textit{F}. However, it is also possible to design the tuple $(B_\infty, C_\infty)$ on each device such that either device~\textit{E} or \textit{F} can directly discover its opposite, which we study in this section.

This form of unidirectional discovery is realized using beacon sequences $B \in B_\infty$, in which the beacons on one device are sent with a fixed temporal relation to the reception windows on the same device. For example, let beacon $b_{F,1}$ on device~\textit{F} be sent by $\zeta$ time-units after reception window $X_{F,1}$, as depicted in Figure~\ref{fig:crossCorrelation}. Further, let such a relation exist on both devices and in every period $T_C$ of the reception window sequence. As previously explained $b_{F,1}$ has a random offset of $\Phi_{F,1}$ time-units from the coordinate origin of device~\textit{E}. The temporal correlation between $B_\infty$ and $C_\infty$ on every device implies that the offset $\Phi_{E,1}$ beacon $b_{E,1}$ has from the coordinate origin of device~\textit{F} is fully determined by $\Phi_{F,1}$ (cf. Figure~\ref{fig:crossCorrelation}). It is:
\begin{equation}
\label{eq:crossCorrelation}
\Phi_{E,1} = \zeta + (\zeta - \Phi_{F,1}) = 2 \cdot \zeta - \Phi_{F,1}
\end{equation}
\nomenclature{$\zeta$}{Fixed temporal distance of a certain beacon from a certain reception window on the same device in every period $T_C$}

By exploiting this temporal relation, a quadruple of sequences $(C_{F,\infty},B_{E,\infty}, C_{E,\infty}, B_{F,\infty})$ can guarantee deterministic one-way discovery, even if the pair $(C_{E,\infty}, B_{F,\infty})$ only covers half of the offsets $\Phi_{F,1} \in [0, T_C]$, by having the pair $(C_{F,\infty}, B_{E,\infty})$ covering the remaining ones. Thereby, the number of beacons that need to be sent per device for guaranteeing one-way discovery can be halved.
\begin{figure}[tb]
	\centering
	\includegraphics[width=\imgWidth]{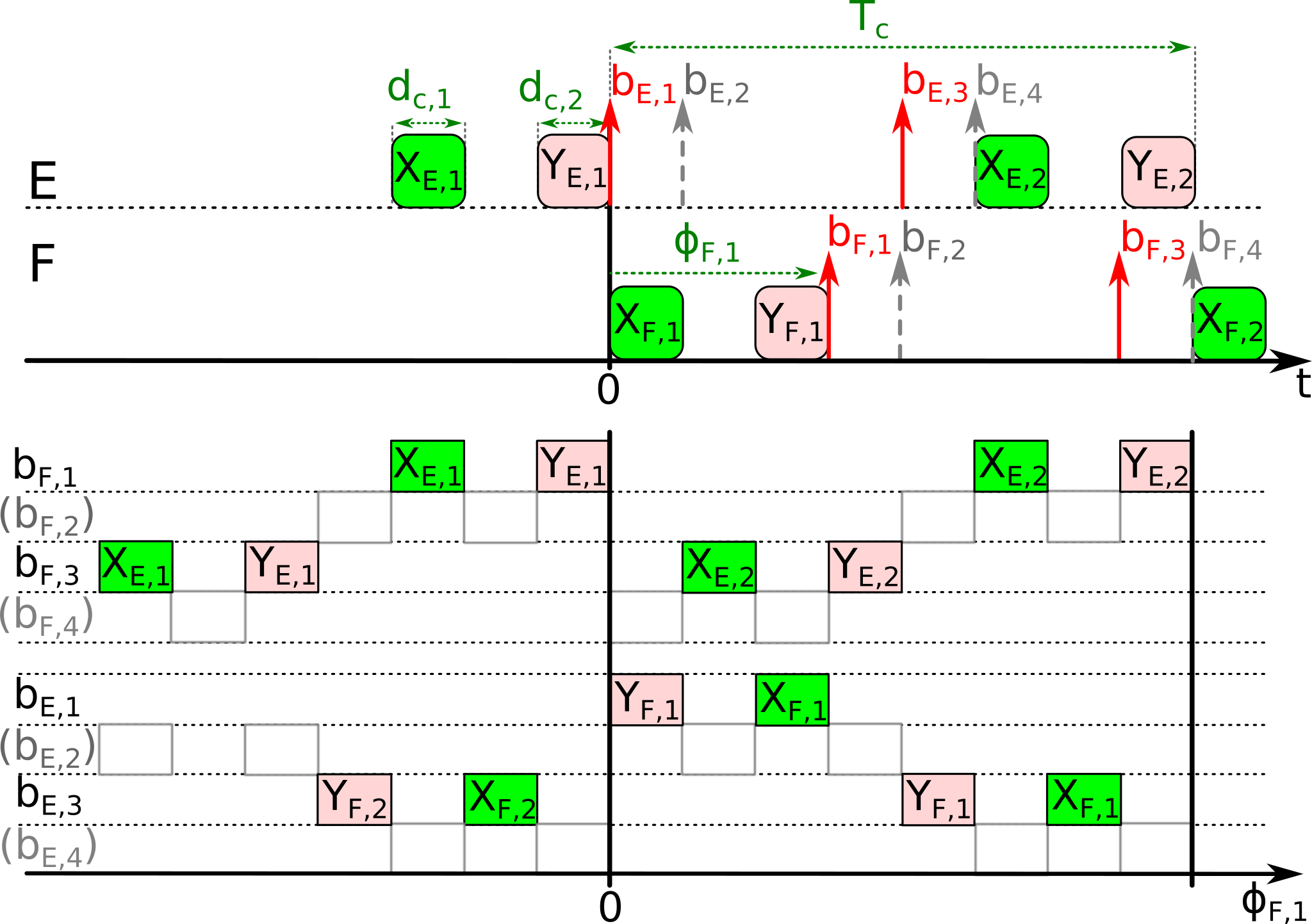}
	\caption{Quadruple of sequences $\mathbf{(C_{E,\infty}, B_{E,\infty},}$ $\mathbf{C_{F,\infty}, B_{F,\infty})}$ that exploits temporal correlations.
	}
	\label{fig:crossCorrelatedSchedule} 
\end{figure}
The upper part of Figure~\ref{fig:crossCorrelatedSchedule} depicts the beacons (arrows) and reception windows (rectangles) of two devices \textit{E} and \textit{F}. On each device, the reception windows and beacons have a fixed temporal relation, whereas beacon $b_{F,1}$ has a random offset $\Phi_{F,1}$ from the coordinate origin of device~\textit{E}. Dashed arrows depict beacons that would need to be sent if every device would have to cover all offsets in the entire period $T_C$ on its own. When exploiting temporal correlations between $B_\infty$ and $C_\infty$ on the same device, these beacons can be omitted without increasing the one-way worst-case latency. The lower part of Figure~\ref{fig:crossCorrelatedSchedule} depicts the coverage map of the beacons $b_{F,1},...,b_{F,4}$ of device~\textit{F} and $b_{E,1},...,b_{E,4}$ of device~\textit{E}. 
This coverage map represents all offsets $\Phi_{F,1}$, for which either a beacon from device~\textit{F} overlaps with a reception window of device~\textit{E} or a beacon from device~\textit{E} overlaps with a reception window of device~\textit{F}.
Covered offsets of omitted beacons have been left white. As can be seen, every possible initial offset $\Phi_{F,1}$ is covered by either a beacon of $B_{F,\infty}$ falling into a reception window of $C_{E,\infty}$, or a beacon of $B_{E,\infty}$ falling into a reception window of $C_{F,\infty}$, and hence the number of beacons per device is halved compared to direct bi-directional discovery. This leads to the following latency bound, which is lower than the one given by Theorem~\ref{thm:boundSymUncorrelated}. Since there are no further possibilities to improve pairwise discovery, this is also the tightest fundamental bound for all pairwise deterministic \ac{ND} protocols.

\begin{theorem}
\label{thm:boundCorrelated}

The lowest worst-case latency a pair of devices can guarantee for mutual exclusive one-way discovery  (i.e., either of both devices can discover its opposite one) is given by:
\begin{equation}
\label{eq:boundCorrelated}
L = \min\left(\left\lceil \frac{1}{\eta} \right\rceil^2 \cdot \frac{\omega \alpha}{\eta \cdot \lceil \nicefrac{1}{\eta} \rceil  - 1/2}, \left\lfloor \frac{1}{\eta} \right\rfloor^2 \cdot \frac{\omega \alpha}{\eta \cdot \lfloor \nicefrac{1}{\eta} \rfloor - 1/2}\right)
\end{equation}
\begin{prf}
For a given set of offsets $\Omega_{F}$ covered by $B_{F} \in B_{F,\infty}$ on device~\textit{E}, Equation \ref{eq:crossCorrelation} defines a set of offsets $\Omega_{E}$ that are automatically covered by $B_{E} \in B_{E,\infty}$ on device $F$, and vice-versa. 
If $\Omega_F$ and $\Omega_E$ are disjoint, the amount of offsets contained in $\Omega_F \cup \Omega_E$ must sum up to $T_C$ time-units for guaranteeing one-way discovery. Hence, the beacon sequence on every device needs to cover only $\nicefrac{1}{2} \cdot T_C$ time-units to guarantee one-way determinism, and Equation~\ref{eq:wc_latency_theorem} becomes:

\begin{equation}
L = \left \lceil \frac{T_C}{2 \cdot \sum_{k=1}^{n_C} d_{k}} \right \rceil \frac{\omega}{\beta}
\end{equation}
The rest of this proof is identical to the one for direct symmetric discovery (cf. Theorem \ref{thm:boundSymUncorrelated}).
\end{prf}
\end{theorem}
Theorem~\ref{thm:boundCorrelated} is valid for one-way discovery (i.e., device~\textit{E} discovers device~\textit{F} \textbf{or} vice-versa). An indirect reverse discovery can be realized as follows. Each device transmits its next point in time at which it listens to the channel along with its beacons. The receiving device then schedules an additional beacon at the received point in time. This technique is called \textit{mutual assistance}~\cite{kindt:17a}, and is actually a form of synchronous connectivity. Here, the latency for two-way discovery will be increased by the maximum temporal distance between any beacon and its succeeding reception window on the same device. An upper bound for this penalty for two-way discovery is $T_C$ time units, which can be reduced significantly in sequences with more than one reception window per period $T_C$.
%


		\section{Table of Symbols}
		\printnomenclature[1.2cm]
		\section{List of Acronyms}
\begin{acronym}
	\acro{ND}[ND]{neighbor discovery}
	\acro{MANET}[MANET]{mobile ad-hoc network}
	\acro{BLE}[BLE]{Bluetooth Low Energy}
	\acro{PI}[PI]{periodic interval}

\end{acronym}
	
	\end{appendix}
\end{document}